\documentclass[prx,twocolumn,english,floatfix,longbibliography]{revtex4-2}
\usepackage[utf8]{inputenc}
\usepackage{booktabs} 
\usepackage{tikz}
\usepackage{graphicx}
\usepackage{amsfonts}
\usepackage{amsmath}
\usepackage{amssymb,bbold}
\usepackage{color}
\usepackage{bm}
\usepackage{bbm}
\usepackage{dsfont}
\usepackage{enumerate}
\usepackage{amsthm} 
\usepackage{mathtools}
\usepackage{array}
\usepackage{mathdots}
\usepackage[colorlinks,bookmarks=false,citecolor=blue,linkcolor=red,urlcolor=blue]{hyperref}
\usepackage{physics}
\usepackage{comment}
\usepackage{svg}
\usepackage[T1]{fontenc}
\DeclareUnicodeCharacter{0097}{\textemdash{}}
\renewcommand{\selectlanguage}[1]{}

\begin{document}

\title{
Causality, localization, and universality\\ of monitored quantum walks with long-range hopping}
\author{Sayan Roy$^1$, Shamik Gupta$^2$, and Giovanna Morigi$^1$}
\affiliation{$^1$Theoretische Physik, Universit\"{a}t des Saarlandes, D-66123 Saarbr\"{u}cken, Germany\\ $^2$Department of Theoretical Physics, Tata Institute of Fundamental Research, Homi Bhabha Road, Mumbai 400005, India}

\begin{abstract}
A powerful strategy to accelerate quantum-walk-based search algorithms leverages on resetting protocols, where a detector monitors a target site and the evolution of the walker is restarted if no detection occurs within a fixed time interval. The optimal resetting rate can be extracted from the time evolution of the probability $S(t)$ that the detector has not clicked up to time $t$. We analyze $S(t)$ for a quantum walk on a one-dimensional lattice when the coupling between sites decays algebraically as $d^{-\alpha}$ with the distance $d$, for $\alpha\in(0,\infty)$. At long times, $S(t)$ decays with a universal power-law exponent that is independent of $\alpha$. At short times, $S(t)$ exhibits a plethora of phase transitions as a function of $\alpha$. From this, we provide a strategy to determine the optimal resetting rate. We identify two regimes: for $\alpha>1$, the resetting rate $r$ is bounded from below by the velocity with which information propagates causally across the lattice; for $\alpha<1$, instead, the long-range hopping tends to localize the walker: The optimal resetting rate depends on the size of the lattice and diverges as $\alpha\to 0$. Our strategy directly connects local measurement outcomes with the global dynamics encoded in $S(t)$. We derive simple models explaining our numerical results, shedding light on the interplay of long-range coherent dynamics, symmetries, and local quantum measurement processes in determining equilibrium. Our findings offer experimentally testable predictions and provide new physical insights on optimizing quantum search through resetting. 
\end{abstract}
\date{\today}
\maketitle

\section{Introduction} 

Efficient search strategies lie at the heart of numerous problems in science and engineering, where the goal is often to locate a stationary target in minimal time. In classical settings, a paradigmatic strategy involves a searcher executing a random walk with intermittent stochastic resetting to its initial position. At each time step, the searcher either undergoes a random displacement or returns to the starting point with a prescribed probability. The performance of such algorithms is typically characterized by the mean first-passage time (MFPT)—the average time taken to reach the target for the first time—whose minimization yields an optimal reset probability enhancing search efficiency~\cite{Evans2011DiffusionResetting,Evans2020StochasticApplications, Gupta2022StochasticReview, nagar2023stochastic}. Such strategies have found applications in diverse setups, from foraging in biological systems~\cite{Benichou2011IntermittentStrategies} to randomized algorithms in computer science~\cite{Luby:1993}, and they have inspired extensive efforts to characterize their nonequilibrium steady states, relaxation timescales, and thermodynamic signatures~\cite{Perfetto2021DesigningResetting,Mukherjee2018QuantumReset,Acharya2023Tight-bindingTimes,Perfetto2022ThermodynamicsResetting,Kulkarni2023GeneratingResetting,Magoni2022EmergentResetting,kulkarni2024dynamically,Turkeshi2022,geher2024reset,wald2024stochastic,Yin:2025}. 

{\it Quantum} walks on a lattice exhibit fundamentally distinct features such as quantum tunneling and interference, which form the basis of several quantum algorithms~\cite{Kempe:2003, Ambainis:2003, Kendon2010}. These dynamics promise a natural quantum advantage for search protocols \cite{Kendon2010}. However, unlike classical random walks, the lack of deterministic trajectories in quantum mechanics complicates the definition of first-passage time \cite{Friedman2017QuantumProblem}. This challenge can be addressed by performing repeated projective measurements at the target site, spaced by a fixed time interval $\tau$, thereby defining the first-detection or hitting time~\cite{Yin2023RestartTimes, Didi2022Measurement-inducedWalks, Yin2019LargeTime, Friedman2017QuantumWalk, Friedman2017QuantumProblem, Thiel2020DarkDetection, Yin2024,Mulken2007SurvivalTrapping, Das2022QuantumChain, Dattagupta:2022, Dubey:2021, Lahiri:2019, Dubey_2023,Kulkarni2023FirstMeasurements,Dhar2015DetectionMeasurements,Dhar2015QuantumModel,Shukla:2025}. The authors of \cite{Yin2023RestartTimes} demonstrate that quantum hitting times on one-dimensional lattices can outperform classical counterparts due to ballistic propagation. Yet, this advantage is limited by dark states, namely, quantum states with zero overlap at the target that evade detection, causing the hitting probability to saturate well below unity. A promising solution involves a sharp resetting protocol~\cite{Yin2023RestartTimes}, where the walker is restarted at the initial site if no detection occurs within a fixed time window. This approach yields an optimal reset rate that minimizes the average first-detection time, establishing a quantum analog of the classical MFPT optimization.

\begin{figure}[!htpb]
    \centering
\includegraphics[width=\columnwidth]{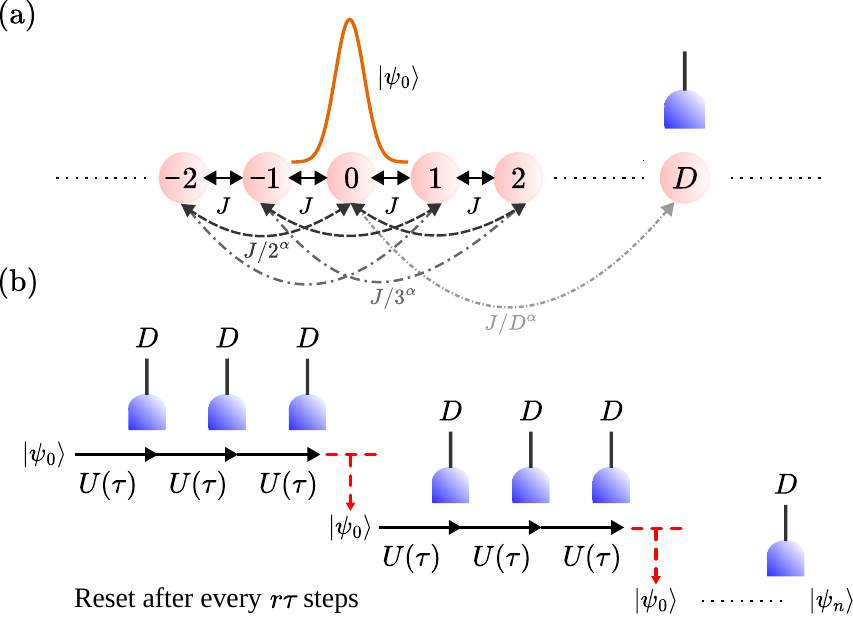}
    \caption{ (a) Monitored quantum walk on a one-dimensional lattice with periodic boundary conditions. The walker is prepared in the site $|0\rangle$, and the target site $|D\rangle$ is monitored by a detector, which clicks when the walker has reached the target. The walker undergoes the dynamics governed by Hamiltonian \eqref{eq:Hamiltonian}, with tunneling amplitude decaying with the distance $D$ as $J/D^\alpha$. Note that in the figure, sites $N-2$ and $N-1$ are labeled $-2$ and $-1$, respectively. (b) Schematic representation of the resetting protocol. The time evolution runs from left to right and alternates unitary evolution for a fixed interval $\tau$ [described by the unitary operator $\hat{U}(\tau)$] with an instantaneous projective measurement by the detector at $D$. The red arrows indicate a resetting event, which occurs in case no click has been recorded in the interval of time $t_{\mathrm{R}} = r\tau$, with $r$ the resetting rate.} \label{fig:scheme}
\end{figure}

Recent advances in quantum simulation platforms, ranging from trapped ions to Rydberg atoms and polar molecules, have unlocked the ability to engineer long-range interactions with controllable power-law decay~\cite{Barredo:2015,Baier:2016,Periwal:2021,Kotibhaskar:2024}. These interactions interpolate between nearest-neighbor and all-to-all coupling regimes, posing a profound question: How does the range of coherent hopping impact the speed and success of a quantum search process? Specifically, one may consider the set-up illustrated in Fig.\ \ref{fig:scheme}(a), in which a particle undergoes coherent hopping to distant sites with amplitudes that decay as a power law in distance, $|x|^{-\alpha}$ with exponent $\alpha > 0$. The detection dynamics in such systems is expected to depend sensitively on $\alpha$, interpolating between the short-range regime considered in Ref.~\cite{Yin2023RestartTimes} and the long-range regime.

A fundamental constraint on the speed of quantum search is provided by the Lieb-Robinson (LR) bound, which limits the spread of quantum correlations and sets a lower bound on the timescale $\tau_{LR}$ over which a quantum walker can reach a target site~\cite{Lieb1972}. This bound depends critically on the structure of the tunneling matrix: for nearest-neighbor hopping, the Lieb-Robinson velocity is constant, enforcing causal, light-cone-like propagation and yielding a linear scaling of $\tau_{LR}$ with the distance $D$ between initial and target sites.
The picture changes dramatically in the presence of long-range tunneling, where hopping amplitudes decay as a power law $1/D^\alpha$. In one dimension, for $\alpha < 2$, the Lieb-Robinson timescale becomes sublinear in $D$, and for $\alpha < 1$ it depends solely on the system size~\cite{Tran:2020,Tran2021Lieb-RobinsonInteractions,Guo:2020}.

 While this might suggest a potential speedup in propagation, long-range hopping can also localize the walker near its initial site, effectively freezing the dynamics~\cite{Santos:2016,Celardo:2016}. In classical settings, search processes with long-range hopping modeled by Lévy flights with jump lengths drawn from heavy-tailed distributions have been shown to exhibit nontrivial interplay with stochastic resetting. Specifically, the mean first-passage time to a target becomes strongly dependent on the Lévy index, with optimal resetting rates emerging only within certain regimes of the power-law exponent~\cite{PhysRevE.92.052127}. These studies reveal that the effectiveness of resetting as a search-enhancement strategy can be significantly altered by the underlying transport mechanism. In light of this, it is natural to inquire whether similar behavior arises in quantum systems featuring long-range coherent hopping. The dependence of detection efficiency on the exponent $\alpha$ in the presence of quantum resetting remains largely unexplored and constitutes the central focus of this work.

 In this work, the central question we address is whether the inherent non-causality of long-range hopping can accelerate or even dramatically boost the convergence time of quantum search protocols. To investigate this, we consider a quantum walk on a one-dimensional lattice with hopping amplitudes scaling as $1/D^\alpha$, as depicted in Fig.\ref{fig:scheme}(a). The target site at distance $D$ is continuously monitored, and the walker is reset to the initial position if no detection occurs within a fixed time interval, as illustrated in Fig.\ref{fig:scheme}(b). We show that the exponent $\alpha$ fundamentally governs the equilibration time and controls the onset of spectral phase transitions. Our results provide a physical understanding of how local measurements can influence relaxation, a global dynamical property, in systems with long-range hopping. Furthermore, they enable us to analytically estimate the optimal resetting rate for the search protocol. Counterintuitively, we demonstrate that long-range hopping induces localization effects that hinder search efficiency, making it less advantageous than nearest-neighbor hopping, which remains optimal for minimizing search times. These findings not only advance the theoretical framework but also open prospects for generating new ideas in controlling quantum dynamics through measurement protocols.
 
This manuscript is organized as follows. In Sec.~\ref{Sec:Model}, we discuss the basic model of a monitored quantum walk with long-range hopping. We then derive the probability that the detector at the target site clicks, i.e., the walker has reached the target site. Starting from this result, in Sec.~\ref{Sec:Timescales}, we discuss the characteristic timescales of the probability of a click as a function of $\alpha$. We discuss the optimal resetting rate in Sec.~\ref{Sec:QSR}. Conclusions are drawn in Sec.~\ref{Sec:Conclusions}. Technical details are relegated to the Appendixes.

\section{Quantum walk on a lattice with long-range hopping}
\label{Sec:Model}

In this section, we introduce the quantum resetting protocol and the key quantities that determine the time needed by the walker to reach a target site. 
Our model is illustrated in Fig.~\ref{fig:scheme}(a): A single walker is constrained to move along a one-dimensional lattice composed of $N$ sites. The unitary evolution of the walker is dictated by the Hamiltonian $\hat{H}$ given by
\begin{align}
    \hat{H} = - J\sum_{i \neq j} \frac{1}{d_{ij}^\alpha} \left(\ket{i}\bra{j} + \text{H.c.} \right);~\alpha>0\,,
    \label{eq:Hamiltonian}
\end{align}
with H.c.~denoting Hermitian conjugate and $|i\rangle$ denoting the state of the walker on site $i$ ($i=0,\ldots,N-1$). The parameters entering the Hamiltonian are
the constant $J>0$, scaling the energy, the dimensionless distance
$d_{ij}$ between sites $i$ and $j$, and the exponent $\alpha>0$. The latter characterizes the decay of the tunneling amplitude with distance and is central to our analysis. For this choice, $J$ is the largest element of the tunneling   matrix. In this work, we assume periodic boundary conditions and $d_{ij}\equiv \mathrm{min}(|i-j|,N-|i-j|)$ is the minimum distance between lattice sites $i$ and $j$.

\subsection{Monitored quantum walk and first-hitting time}
\label{sec:St}
Since quantum systems lack well-defined trajectories, the first-hitting time is defined through repeated monitoring at the target site $D$: The detector will reveal the presence of the walker at the target site \cite{Varbanov:2008}. In this work we assume that the monitoring is stroboscopic: the projective measurements are performed at 
times $t_\ell=\ell\tau$, with $\ell=1,2,\ldots$, and separated by the finite interval $\tau>0$. The scheme is illustrated in Fig.~\ref{fig:scheme}(b): Starting from an initial state $\ket{\psi_0}$ localized at one lattice site, the monitored dynamics alternates unitary evolution and projective measurements. The unitary evolution over the fixed time $\tau$ is governed by the unitary operator 
\begin{equation}
\hat{U}(\tau)= \exp(-{\rm i}\hat Ht/\hbar)\,,
\end{equation} 
where $\hat H$ is the Hamiltonian of the quantum walk given by Eq.\ \eqref{eq:Hamiltonian}. The probabilistic nature of detection allows one to envisage a typical run of the dynamics, which leads to a string of binary measurement outcomes \textit{no} (the walker has not been detected), \textit{no, no}, $\ldots$, ending with a \textit{yes} (the walker has been detected) at, say, time $t_n=n\tau$, namely, the $n$th time step or the $n$th detection attempt.  One then defines $t_n$ as the first-hitting time for the  dynamical run under consideration. Let $F(t_n)$ denote the probability to first detect the walker at the $n$th attempt. The detection probability up to the time $t_n$ is clearly given by the sum of the detection probabilities at each instant $t_l=l\tau$ with $1\le l\le n$:
\begin{eqnarray*}
P_\mathrm{det}(t_n)=\sum_{l=1}^n F(t_l)\,.
\end{eqnarray*}
The corresponding probability that the detector has {\it not} clicked (detected) is 
$$S(t_n)=1-P_\mathrm{det}(t_n)\,.$$ 
In what follows we will refer to it as the ``survival probability,'' using the language used in the context of quantum stochastic resetting. Correspondingly, the first detection probability at time $t_n$ is the difference between the survival probabilities at consecutive times: $F(t_n)=S(t_{n-1})-S(t_n)$. The resetting protocol considered here imposes a maximal time $t_{\rm{R}}=r\tau$, such that if for a given dynamical run no click occurs within $t_{\rm R}$, the walker is instantaneously reset to the initial site and the dynamics starts again. For nearest-neighbor hopping, this strategy provably leads to faster convergence than ballistic propagation \cite{Yin2023RestartTimes,Yin2024}. This is also the strategy of randomized algorithms \cite{Luby:1993}.

As in randomized algorithms, our goal is to identify the optimal resetting rate $r$ leading to the shortest timescale for detection.  We perform the analysis by studying the dynamics of the survival probability as a function of $\alpha$, since this quantity describes the joint effect of unitary propagation and local measurement and thus contains the information about the first hitting time. For this purpose, it is useful to first review the timescales of the unitary dynamics.

\subsection{Timescales of the unitary dynamics}

We review the timescales of the unitary dynamics centering the discussion on the Lieb-Robinson bound. In fact, since this provides a bound on the timescale that correlations are established between two distant sites, it will also influence the probability that a detector, monitoring the target site, clicks as a function of time. The limit $\alpha\to\infty$ in Eq.\ \eqref{eq:Hamiltonian} models tunneling that couples nearest-neighbor sites. When the walker is initially localized at a single site, the minimal timescale $\tau_{\rm LR}$ needed by the walker to reach the target site depends linearly on the distance $D$ and is estimated using the Lieb-Robinson velocity $v_{\rm LR}$ as that for the ballistic motion of a pointlike particle, $\tau_{\rm LR}\sim D/v_{\rm LR}$ \cite{Lieb1972,Tran:2020}. The Lieb-Robinson velocity $v_{\rm LR}$, in turn, is proportional to $J$. Due to the light-cone-like causality, at times $t<\tau_{\rm LR}$, the probability that the walker has reached the target is negligible. Correspondingly, a detector that locally monitors the target site will not click. This consideration suggests that the resetting rate shall be bound from below by the Lieb-Robinson velocity. We expect, in particular, that a faster resetting rate, below this bound, will tend to localize the walker about the initial site, hindering it to reach the target. A question we will address in this paper is whether the Lieb-Robinson velocity also determines the optimal resetting rate.

We try now to follow this reasoning for finite value of $\alpha$. For this purpose, in Table \ref{Table:1} we summarize the bounds determined in the literature~\cite{Tran:2020,Tran2021Lieb-RobinsonInteractions,Guo:2020}. It turns out that the previous consideration can be extended to finite values of $\alpha$, as long as $\alpha\gtrsim 2$. For $\alpha<2$, 
it is noteworthy that the quantity $\tau_\mathrm{LR}$ exhibits three distinctive change in behavior at $\alpha=2$, $\alpha=1$ and $\alpha=1/2$. In fact, the bound scales sublinearly with the distance for $1<\alpha<2$, it becomes independent of the distance for $\alpha<1$, and it depends solely on the total number of lattice sites for $\alpha<1/2$. In the latter regime, by increasing the total number of lattice sites $N$, the minimal time $\tau_{\rm LR}$ approaches zero, suggesting that the walker reaches the target almost instantaneously.

\setlength{\tabcolsep}{16pt}
\begin{table}[h]
\centering
\renewcommand{\arraystretch}{1.4}  
\begin{tabular}{@{}c c  c@{}}
\toprule
\toprule
Exponent $\alpha$ &  Lieb-Robinson time $\tau_{\rm LR}$ & Ref.\\
\midrule
$(2, \infty)$ & $O(D)$ & \cite{Tran:2020} \\
$(1,2)$ & $O(D^{\alpha - 1})$& \cite{Tran:2020}\\
$1$ & $O(\log D)$ &\cite{Tran:2020} \\
$(1/2, 1)$ & $O(1)$ &\cite{Tran:2020} \\
$(0,1/2)$ & $O(N^{\alpha - 1/2})$ &\cite{Guo:2020}\\
\bottomrule
\bottomrule
\end{tabular}
\caption{\label{Table:1} The lowest bound on the timescale with which information propagates over a distance $D$ according to the dynamics of Eq.\ \eqref{eq:Hamiltonian} and as a function of $\alpha$.
The Lieb-Robinson time $\tau_{\rm LR}$ refers to the minimal time required for quantum state transfer from an initially localized state at a single lattice site to another site at a distance $D$. $N$ is the total number of lattice sites. The ``big-$O$'' notation $f(x) = O(g(x))$ indicates the existence of constants $c_1$ and $N_1$ such that $0 \leq f (x) \leq  c_1\, g(x)$ for all $x \geq N_1$. The column ``Ref.'' provides the reference to the work where the bound has been reported.}
\end{table}

We note that the scalings in Table \ref{Table:1} would change if one scales the constant $J$ in Eq.~\eqref{eq:Hamiltonian} by a size-dependent factor $\widetilde{N} \equiv \sum_{j=1}^N d_{ij}^{-\alpha}~\forall~i$. This procedure is called Kac scaling~\cite{campa2014physics,alpha-kura-shamik} and it is customary
while studying the thermodynamic properties of long-range Hamiltonians to make the energy of the system extensive as $N \to \infty$. In our analysis, however, any such scaling leads eventually to a mere rescaling of the hopping factor $J$. We refrain from applying it therefore, as we are interested in determining the timescale of the quantum resetting protocol comparing the dynamics at fixed lattice size $N$ by solely varying $\alpha$.

\subsection{Detection and survival probability}

We now come to the specifics of the dynamics with unitary evolution interspersed with projective measurements. Let the initial state of the walker be $\ket{\psi_0}=|0\rangle$ and the target state be $|D\rangle$ with $D \ne 0$, i.e., the initial state of the walker could be any site other than the detector site itself. 
The probability of a detection at time $t_1=\tau$ is given by the expectation value of the projector $|D\rangle\langle D|$ on the state $\ket{\psi_1}=U(\tau)|\psi_0\rangle$, namely, $P_{\rm det}(t_1)=\langle \psi_1|D\rangle\langle D|\psi_1\rangle$. In this case, the protocol has been successful and stops. Assume now no detection. The probability that the detector does not click (the survival probability) is $S(t_1)=1-P_{\rm det}(t_1)$ and the state immediately after the detection is $|\psi_1\rangle=|\psi_1^+\rangle/\sqrt{S(t_1)}$, with
\begin{equation}
\ket{\psi_1^+}=[(\hat{I}_N-|D\rangle\langle D|)\hat{U}(\tau)] |\psi_0\rangle\,,
\end{equation}
and $\hat{I}_N$ is the identity operator. Iterating the procedure, the probability for a series of ``no'' outcomes at times $t_\ell\le t_n$ can be understood as the norm of the unnormalized state
\begin{equation}
|\psi_n^+\rangle =[(\hat{I}_N-|D\rangle\langle D|)\hat{U}(\tau)]^n |\psi_0\rangle \,,
\label{eq:psinplus}
\end{equation}
such that
\begin{equation}
\label{Eq:survival}
S(t_n)=\langle \psi_n^+|\psi_n^+\rangle\,.   
\end{equation} 
The quantum mechanical state immediately after the ``no'' outcome at time $t_n$ is then $|\psi_n\rangle=|\psi_n^+\rangle/\sqrt{S(t_n)}$.

The survival probability is closely related to the delay distribution between detection events in atomic spectroscopy \cite{Cohen-Tannoudji:1986}, which is the cornerstone of the theory of quantum trajectories \cite{Dalibard:1992,Dum1992,Plenio:1998}. It contains the information about the detection statistics, and hence its study will permit us to determine the optimal resetting time of the protocol. 

\begin{figure}[!htpb]
    \centering
\includegraphics[width=\columnwidth]{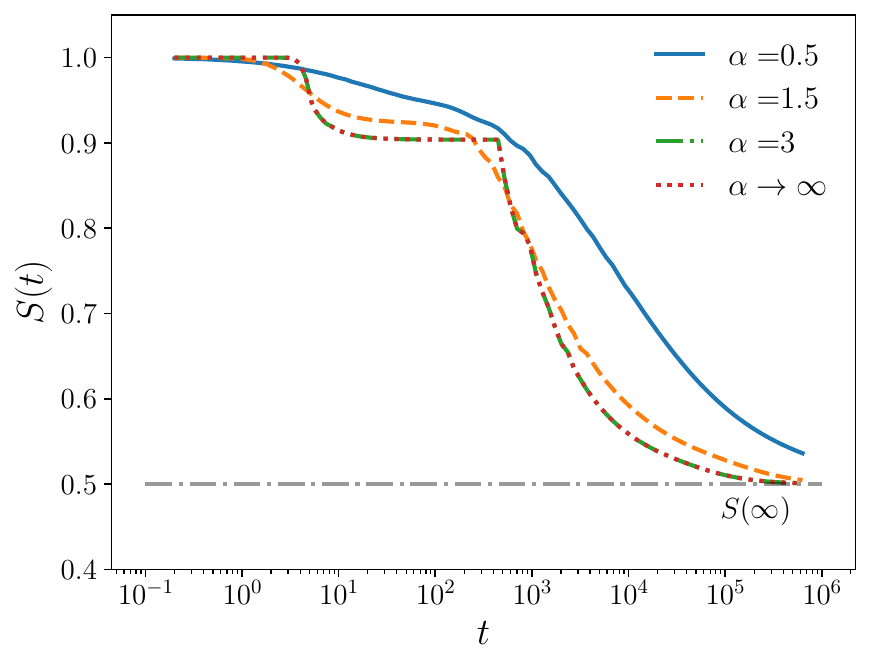}
    \caption{Time evolution of the survival probability, Eq.\ \eqref{Eq:survival}, for four different values of the exponent $\alpha$. The parameters are $N = 1000, \tau = 0.2/J, D = 10$, and $J=1$. The gray dot-dashed horizontal line indicates the asymptotic value of the survival probability, denoted by $S(\infty)$.}
    \label{fig:sprob_slice}
\end{figure}

In Fig.~\ref{fig:sprob_slice} the dynamics of the survival probability is displayed for different values of $\alpha$, representative for the different regimes of the Lieb-Robinson bound according to Table \ref{Table:1}: $\alpha=3$ and $\alpha\to\infty$ (nearest-neighbor) for the linear scaling with the distance from the target, $\alpha=3/2$ for the sublinear scaling, and $\alpha=1/2$ for the regime where the bound is independent of the distance. A first observation is that at the asymptotics, the survival probability $S(t)$ saturates at $0.5$ for all values of $\alpha$, namely, $S(\infty)\equiv\lim_{t\to\infty} S(t)=0.5$. At finite time, the time evolution exhibits different features depending on $\alpha$. While for $\alpha>2$ we can identify three well-distinguished timescales, which we will associate with the causal propagation of information (discussed in detail in Sec.~\ref{sec:regime2}), for $1\lesssim \alpha \lesssim 2$ this behavior smoothes out, and almost disappears for $\alpha <1$.

 \begin{figure*}[!htpb]
 \centering
\includegraphics[scale =0.35]{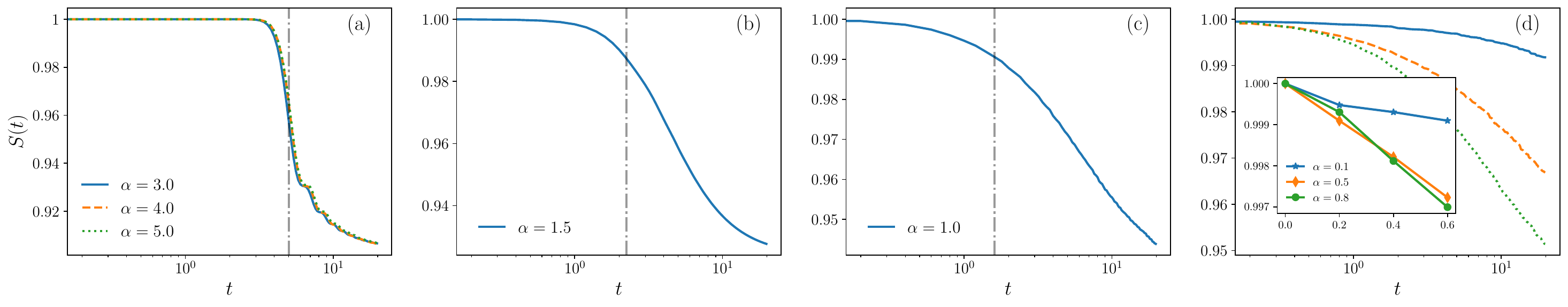}
\caption{Dynamics of the survival probability $S(t)$ at short times and for different values of $\alpha$. Subplot (a) shows the dynamics for three values $\alpha>2$, where causal propagation is expected; (b) displays $\alpha=3/2$, where the Lieb-Robinson bound scales sublinearly; (c) shows $\alpha=1.0$, where the scaling is logarithmic, while in panel (d) $\alpha<1$. The inset in (d) zooms on the short time behavior for the first three stroboscopic measurements. In all plots, the gray dot-dashed vertical lines indicate the timescale $\tau_{\rm LR}$ as given in Table~\ref{Table:1}. The parameter values are $N = 1000, \tau = 0.2/J, J=1, D = 10$. Time is in units of $1/J$.}
    \label{fig:LR}
\end{figure*}

Figure \ref{fig:LR} displays the short-time dynamics of the survival probability $S(t)$, Eq.~\eqref{Eq:survival}, for more values of the exponent $\alpha$. For $\alpha>2$, panel (a), the survival probability is practically unity until a timescale of the order of the Lieb-Robinson time, after which it quickly decays. This behavior agrees with the picture of causal propagation of the wave packet: For $t<\tau_{\rm LR}$, the wave packet has not reached the target, and the detector does not click [$S(t)=1$]. For $t\gtrsim \tau_{\rm LR}$, instead, the detection probability quickly grows and $S(t)$ quickly decreases. For $\alpha$ below 2, this sharp separation into two regimes gets gradually lost. For $0.5<\alpha<2$, panels (b) and (c), the probability of a click at short times increases, while at larger times, it decays more slowly. For $\alpha<0.5$, panel (d), the decay of the survival probability becomes increasingly slower as $\alpha$ is decreased.  Yet, differing from the short-range dynamics, there is a small but finite detection probability at $t_1=\tau$, as visible in the inset.

\begin{figure}[!htpb]
    \centering
\includegraphics[width=\columnwidth]{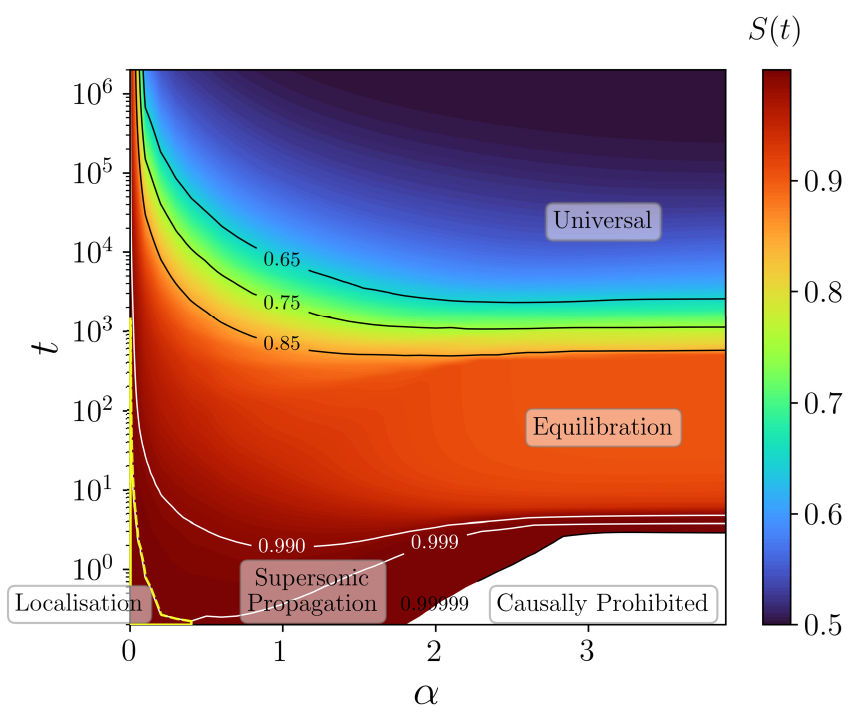}
    \caption{Density plot of the survival probability $S(t)$, Eq.\ \eqref{eq:psinplus}, in the $(\alpha,t)$-plane. Everywhere in the white region, the survival probability value equals unity [no decay from its initial value $S(0)=1$]. The black lines are the equipotential lines, with the corresponding value of $S(t)$ explicitly given.  The parameter values are $N = 1000, \tau = 0.2/J, J=1, D = 10$. The time is in units of $1/J$.}
    \label{fig:phase_Diagram}
\end{figure}

Figure \ref{fig:phase_Diagram} summarizes the salient features of the dynamics of the survival probability as a density plot in the $t-\alpha$ plane. The color code allows for the identification of different regimes as a function of time. The white region corresponds to $S(t)=1$, and thus the probability that the detector clicks is zero. This is strictly found only for $\alpha\gtrsim 2$. The dark red region indicates values of the survival probability that are close to, but still smaller than, unity. In this region there is an infinitesimal yet finite probability that the detector clicks. The color scale goes through an intermediate region and finally reaches the blue region, where $S(t)$ decays algebraically with time with a functional behavior that is independent of $\alpha$. The equipotential lines correspond to fixed values of $S(t)$, and they show that, while the above-mentioned regimes are generally present for all exponents $\alpha$, the characteristic timescales do depend on the exponent, even differing by orders of magnitude. The first equipotential line and the nonmonotonic behavior of the second equipotential line, in particular, turn out to separate dynamics with a very different physical origin.

The analysis in the next section leads us to argue that the short time dynamics is separated into three regimes: (i) The white region for $\alpha>2$ is where the detection probability is exactly zero and the survival probability is unity for a finite interval of time. This picture invokes a concept of causality, corresponding to which the walker propagates at a finite velocity and has not yet reached the target site, and in this sense, we denote it as the ``causally prohibited'' region. (ii) The region for $\alpha<2$ is where there is an infinitesimal yet finite probability that the detector clicks even at the first measurement. We denote it as the ``supersonic propagation'' regime, since the walker propagates faster than the linear light cone. (iii) Within this region, the second equipotential line for $\alpha\lesssim 1/2$ grows to increasingly large times as $\alpha$ tends to zero, Here, we will show that the dynamics is essentially frozen out due to the long-range diffusion. This localization phenomenon is due to the flat dispersion spectrum of the Hamiltonian and is analogous to what has been called ``cooperative shielding'' in Ref.\ \cite{Santos:2016,Celardo:2016}. Using this terminology, we label this region  ``localization.'' 

After the initial dynamics, the survival probability exhibits (iv) an intermediate region. This region is characterized by transient features that emerge from interference between wave packets propagating along both directions of the lattice. Here, finite-size effects become relevant, and the dynamics reaches a stationary state. Based on these observations, we denote it as the ``equilibration'' regime. (v) In the asymptotic regime, the survival probability decays algebraically with time to the asymptotic value $S(\infty)=1/2$ with an exponent that is independent of $\alpha$. For this reason, we denote this regime as ``universal.''

These physical insights are detailed in the next section and are the key to identify the optimal resetting rate speeding up the convergence to the target site, as we will argue in Sec.\ \ref{Sec:QSR}.

\section{Timescales of monitored quantum walk}
\label{Sec:Timescales}

In this section, we provide insights into the behavior of the survival probability, identifying the characteristic timescales and arguing about their physical origin. We focus on a specific regime in which the unitary evolution between two detection events weakly perturbs the state of the walker, corresponding to 
$J\tau\ll 1$. In this regime, the dynamics of Eq.\ \eqref{eq:psinplus}, alternating coherent evolution with detection events, can be cast in the form of a continuous dynamics,
\begin{align}\ket{\psi_n^+}= \hat{\tilde{U}}(t_n)\ket{\psi_0}\,,
\label{eq:S_n}
\end{align}
governed by a non-unitary evolution operator 
\begin{align}
    {\hat{\tilde{U}}}(t)
    &\approx e^{-i\hat{H}_\mathrm{eff}t/\hbar}\,,
    \label{eq:U-eff}
\end{align}
where $\hat H_{\rm eff}$ is a non-Hermitian Hamiltonian, which we derive in the next subsection. This formulation forms the cornerstone of our theoretical understanding and allows us to single out the universal and characteristic features of the detection probability. 
 
\subsection{Equivalent non-Hermitian quantum walk}

The Hilbert space $\mathcal{H}$ of the model~\eqref{eq:Hamiltonian} is subject to repeated measurements projecting onto site $D$. It is therefore conveniently decomposed in terms of the detector Hilbert space $\mathcal{H}_D$, consisting of state $|D\rangle$, and its complementary Hilbert space $\mathcal{H}^c_D$, consisting of the rest of the lattice without the detector site, as $\mathcal{H}=\mathcal{H}_D \oplus \mathcal{H}^c_D$.
We denote by $$\hat{P}_D^c\equiv\hat{I}_N-|D\rangle\langle D|$$ the projection operator onto the subspace $\mathcal{H}^c_D$.
Using this definition, we cast Eq.\ \eqref{eq:psinplus} into the form
\begin{equation}
    \ket{\psi_n^+}=\hat{P}^c_D\hat{U}\ket{\psi_{n-1}^+}\equiv\hat{\tilde{U}}\ket{\psi_{n-1}^+}\,,
\end{equation}
with $\hat{\tilde{U}}\equiv \hat{P}_D^c \hat{U}$ an effective time-evolution operator, which is nonunitary.
With these definitions, the evolution of the unnormalized state $\ket{\psi_n^+}$ may now be written as in Eq.\ \eqref{eq:S_n}.

In the limit of small interval $\tau$ between successive detection attempts, such that $J\tau\ll 1$, we show in Appendix~\ref{app1} that the effective evolution operator, to $O(\tau^2)$, leads to Eq.\ \eqref{eq:U-eff} with the identification of the non-Hermitian Hamiltonian $\hat{H}_\mathrm{eff}$ defined in the Hilbert space $\mathcal H_D^c$:
\begin{equation}
   \hat{H}_\mathrm{eff}\equiv \hat{P}_D^c \hat{H} \hat{P}_D^c - \frac{i\tau}{2} \hat{P}_D^c\hat{H}|D\rangle\langle D|\hat{H}\hat{P}_D^c\,, \label{H_eff}
\end{equation}
where from now on we set $\hbar=1$. Interestingly, $\hat{H}_\mathrm{eff}$ is symmetric, $\hat{H}_\mathrm{eff}=\hat{H}_\mathrm{eff}^T$, with $(\cdot)^T$ denoting the transpose operation. The spectral analysis of $\hat{H}_\mathrm{eff}$ provides valuable information about the behavior of the survival probability $S(t)$. 

Due to its non-Hermiticity, $\hat{H}_\mathrm{eff}$ has a complex spectrum of eigenvalues, which we will denote by $\lambda_a$. It is diagonal in the biorthogonal basis composed of left and right eigenvectors: $\{|\lambda_a\rangle,|\bar\lambda_a\rangle\}$, with $\hat{H}_\mathrm{eff}|\lambda_a\rangle=\lambda_a|\lambda_a\rangle$ and $\langle \bar\lambda_a|\hat{H}_\mathrm{eff}=\lambda_a\langle \bar\lambda_a|$, such that the orthogonality relation holds as $\langle \bar\lambda_a|\lambda_b\rangle=\delta_{a,b}$. Using the completeness relation $\sum_{a=1}^N|\lambda_a\rangle\langle \bar\lambda_a|=\hat{I}_N$, Eq.\ \eqref{eq:S_n} can be cast in the form 
$$\ket{\psi_n^+} = \sum_a e^{-in\tau\lambda_a}\ket{\lambda_a}\bra{\bar\lambda_a}\ket{\psi_0}\,,$$ 
leading to the following exact expression for the survival probability:
\begin{align}
S(t) &=\sum_{a,b} e^{-in\tau (\lambda_a - \lambda_b^*)}\bra{\psi_0} \bar\lambda_b\rangle \bra{\bar\lambda_a}\ket{\psi_0} \bra{\lambda_b}\ket{\lambda_a}\,.
\label{eq:Sn-exact}
\end{align}
The dynamics of non-Hermitian systems can exhibit rich and nontrivial behavior due to the properties of the biorthogonal basis~\cite{Hatano:1996}. In our case, however, the left and right eigenvectors are well approximated by the eigenvectors of the Hermitian part of the effective Hamiltonian~\eqref{H_eff}, namely, of the  operator $\hat{H}_0 \equiv (\hat H_{\rm eff}+\hat H_{\rm eff}^\dagger)/2$. In fact, the non-Hermitian part, $(\hat{H}_{\rm{eff}} - \hat H_{\rm eff}^\dagger)/2$, scales as $J\tau$ and can be treated as a perturbation in the limit of $J\tau \ll 1$~\cite{sakurai, Dhar2015DetectionMeasurements, Dhar2015QuantumModel}. 
Denoting by $|\lambda_a^{(0)}\rangle$ the eigenvectors of the bare Hamiltonian $\hat{H}_0$, in Appendix~\ref{app1:A} we show that 
\begin{align}
S(t) \approx \sum_{a} e^{-2  (\gamma_a\tau) t}|\langle\lambda_a^{(0)}|\psi_0\rangle|^2,
\label{eq:Sn-2}
\end{align}
where we define $\gamma_a=\mathrm{Im}\{\lambda_a\}/\tau$. The equality holds in first-order perturbation theory in the small parameter $J\tau\ll 1$, and now $\gamma_a$ denotes the first-order correction to the real eigenvalue $\lambda_a^{(0)}$ of the bare Hamiltonian $\hat{H}_0$: $\lambda_a=\lambda_a^{(0)}-i\tau \gamma_a$. The imaginary eigenvalues $\gamma_a$ determine the timescales of the decay of $S(t)$ and are therefore the main element of our analysis.

\begin{figure}[!htpb]
    \centering
\includegraphics[width=\columnwidth]{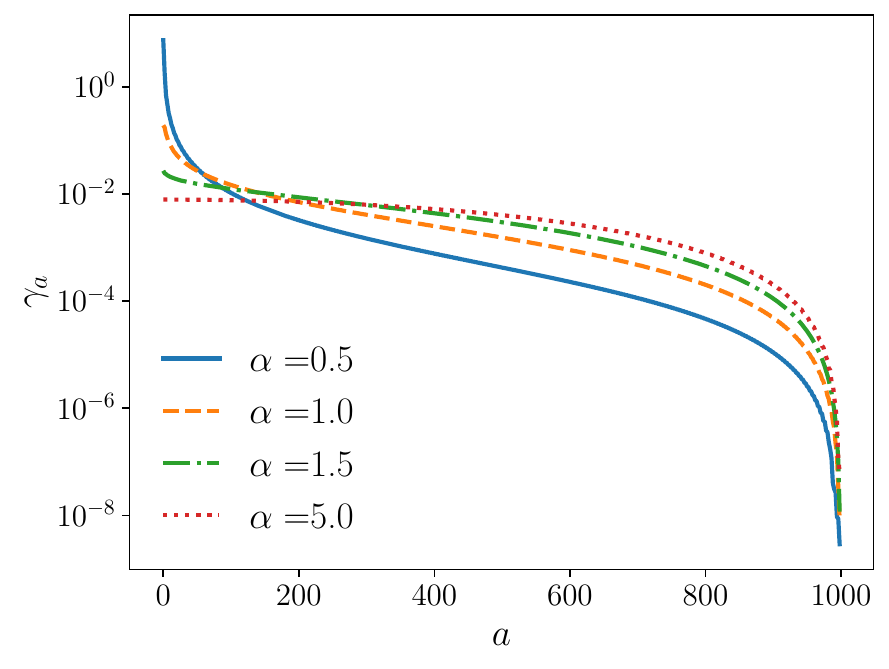}
    \caption{Nonvanishing imaginary part $\gamma_a$ of the eigenvalue spectrum of $\hat H_{\rm eff}$, ordered from the largest to the smallest, for four values of the power-law exponent $\alpha$. The other parameters are $N = 1000$, $\tau = 0.2/J$, $J = 1$, $D = 10$.}
    \label{fig:decay_rate}
\end{figure}

Figure~\ref{fig:decay_rate} displays the nonzero imaginary part $\gamma_a$, ordered from the largest to the smallest value, for four representative exponents $\alpha$. We observe that the interval over which they span increases by orders of magnitude as $\alpha$ decreases below unity. This is exemplified by the behavior of the largest value of $\gamma_a$, which dominates the short-time behavior of the survival probability. On the other hand, the spectrum for $\gamma_a\to 0$ determines the long-time behavior and turns out to be largely independent of $\alpha$. The plot does not show the $N/2$ eigenmodes with $\gamma_a=0$. These eigenmodes have zero spatial overlap with the detector site and are denoted as ``dark eigenmodes.''. As a consequence, any wavepacket consisting of superposition of these mode will never reach the target site and in this sense the eigenmodes are ``dark.'' As we show in Appendix \ref{app2}, they are solely determined by the spatial geometric properties of the setup and are thus independent of the hopping exponent $\alpha$. They are responsible for the asymptotic behavior of the survival probability, $S(\infty)=1/2$.
\begin{figure}[!htpb]
   \centering
\includegraphics[width=\columnwidth]{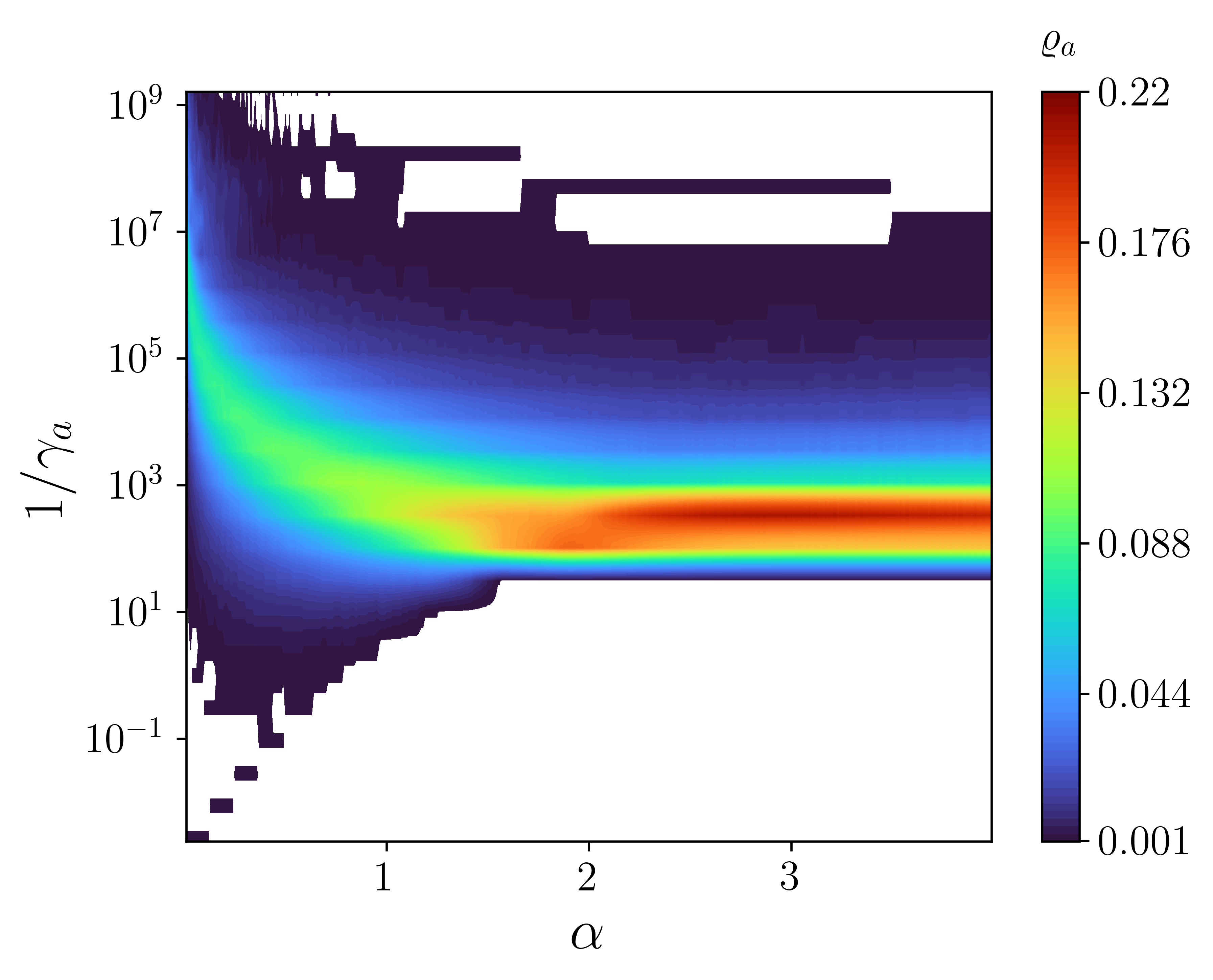}
    \caption{Density plot of the density of modes $\varrho_a=n_a/N$ in the $(\alpha,\gamma_a^{-1})$ plane. Here, $n_a$ is the number of modes with inverse decay rate in the interval $[\gamma_a^{-1}, \gamma_a^{-1} + \delta]$, where $\delta$ is a preassigned small number with the dimension of time. Note that for $\gamma_a = 0$, then $n_a/N=0.5$; see Appendix~\ref{app2}. The parameters are the same as in Fig.~\ref{fig:phase_Diagram}. The rates are in units of $J$.}
    \label{fig:densitystates}
\end{figure}

We can understand the behavior of the survival probability, Fig.\ \ref{fig:phase_Diagram}, from the spectral properties of the non-Hermitian Hamiltonian.  Figure \ref{fig:densitystates} displays the density of states as a function of $\alpha$ and $1/\gamma_a$. 
The density plot allows us to visualize the density of eigenmodes contributing to the short-time behavior (small $1/\gamma_a$), the intermediate behavior, and the long-time behavior (large $1/\gamma_a$). The structure can be put in direct connection with Fig.\ \ref{fig:phase_Diagram} for $\alpha\gtrsim 1/2$ , and in particular follows the same equipotential lines. This relatively simple relation does not apply for $\alpha\lesssim 1/2$: The inverse of the largest eigenvalue goes to zero as $\alpha\to 0$, but is separated by a gap from the rest of the imaginary spectrum. We will show that this gap increases with the size of the lattice $N$ and is responsible for the ``localization'' regime.  Equipped with this model, in the rest of this section we analyze the different regimes individually, and we determine their scaling with the lattice size.

\subsection{The minimal hitting time}

The initial behavior of the survival probability $S(t)$ is determined by the mode with the largest $\gamma_a$ value. Let us denote this value by $\gamma^{\rm max}$, such that $\gamma^{\rm max}\equiv\max_a\{\gamma_a\}$. Based on physical intuition, one would expect that this quantity is directly related to the Lieb-Robinson bound, namely, $v_{\rm LR}\sim \gamma^{\rm max}$. In fact, $\gamma^{\rm max}$ is expected to determine the minimal hitting time, which in turn is related to how fast information can propagate from the initial to the target site. 

\begin{figure}[!htpb]
\includegraphics[width=\columnwidth]{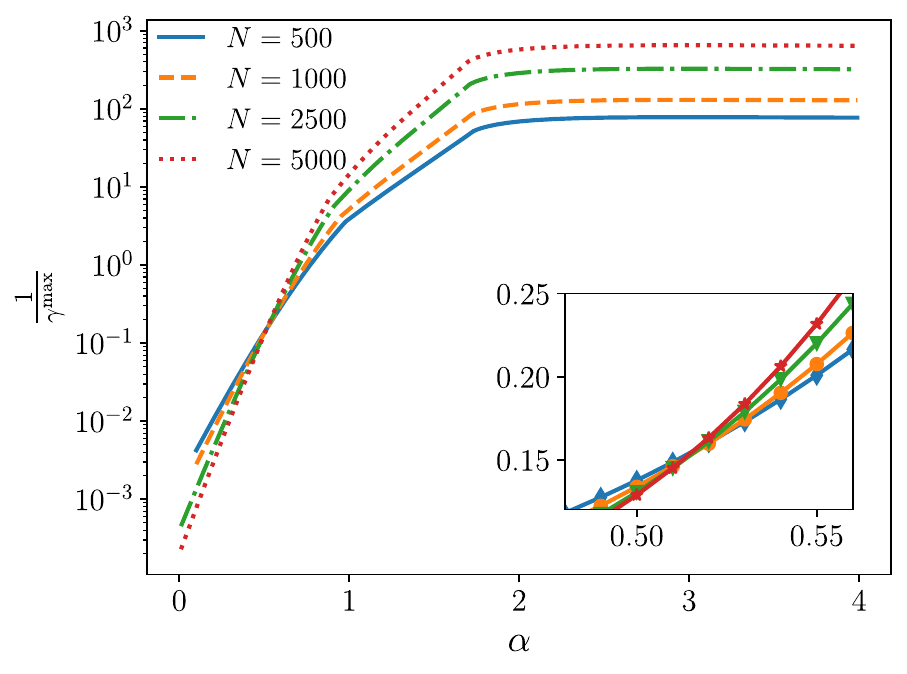}
    \caption{Inverse of the maximal decay rate $\gamma^{\rm max}$ as a function of $\alpha$ for four lattice sizes $N$. The inset zooms into the point where the three curves intersect at $\alpha\sim 1/2$.  Note that $\gamma^{\rm max}$ gives the timescale of the fastest decay of the survival probability $S(t)$. The rates are in units of $J$.}
    \label{fig:gamma_max}
\end{figure}

Figure \ref{fig:gamma_max} illustrates the dependence of the maximal decay rate on $\alpha$ for different lattice sizes $N$.  We observe that it is constant for $\alpha>2$, while instead $\gamma^{\rm max}$ increases as $\alpha$ decreases towards zero. We denote this regime by ``supersonic propagation,'' in contrast to the regime of causal diffusion for $\alpha>2$. Interestingly, the curves at different $N$ intersect at $\alpha=\alpha^* \sim 0.5$. This exponent separates the regime $\alpha\in(\alpha^*,\infty)$ where $\gamma^{\rm max}$ decreases with $N$, from the regime $\alpha\in (0,\alpha^*))$ where it instead exhibits the opposite behavior. We reproduce this behavior by means of a simple model, which approximates the high-energy part of the imaginary spectrum $\gamma_a$; see Appendix \ref{app1:A}. For $0<\alpha<1$, the fast-decaying eigenmodes $|\lambda_a^{(0)}\rangle$ are well approximated by standing waves with wave number $k$ (eigenmodes of $\hat H$) such that 
\begin{equation}
\gamma_a \simeq  \gamma_k = \frac{1}{N}\cos^2(kD)\mathcal J_k^2\,,
\label{eq:gamma:k:main}
\end{equation}
where 
\begin{equation}
\label{eq:spectrum}
    \mathcal J_k=-2J\left(\sum_{r=1}^{N/2}\frac{\cos(kr)}{r^\alpha}\right)
\end{equation}
is the spectrum of Hamiltonian \eqref{eq:Hamiltonian}. The fastest decay corresponds to the mode at $k=0$, with $\gamma^{\rm max}\equiv \gamma_0\sim 2J^2 N^{1-2\alpha}$, and reproduces the behavior of $\gamma^{\rm max}$ in Fig.\ \ref{fig:gamma_max}, with a slight underestimation of $\alpha^*$ (see also Fig.~\ref{fig:gamma_max_approx}). It shows that the behavior of $\gamma^{\rm max}$ is this regime is determined by the modes at wavenumber $k\sim 0$ of the long-range Hamiltonian, which in turn determines the size-dependent scaling of the Lieb-Robinson bound. Equation \eqref{eq:gamma:k:main} thus establishes an explicit connection between the minimal hitting time and $\tau_{\rm LR}$ for $\alpha<1$.

\begin{figure}[!htpb]
\includegraphics[width=\columnwidth]{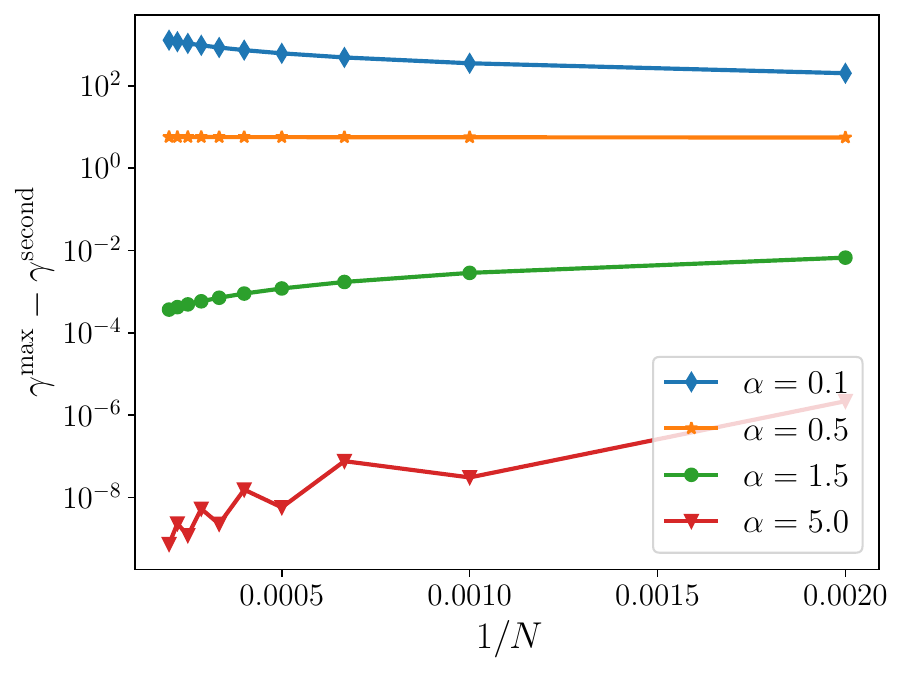}
    \caption{Gap between the largest and the second largest decay rates $\gamma_a$, i.e., $\Delta\gamma=\gamma^{\rm max}-\gamma^{\rm second}$, as a function of $N$ for different values of $\alpha$. The gap is in units of $J$.}
    \label{fig:gap}
\end{figure}

The short-time behavior of the survival probability is also determined by the density of states close to $\gamma^{\rm max}$, see Fig.\ \ref{fig:densitystates}. We expect a clear difference between the dynamics for $\alpha>1/2$, where $\gamma_a$'s form a continuum, from the behavior for $\alpha\lesssim 1/2$, where the eigenvalue $\gamma^{\rm max}$ is separated from the rest of the spectrum by a finite gap. Figure~\ref{fig:gap} displays the imaginary gap between $\gamma^{\rm max}$  and the second largest eigenvalue $\gamma^{\rm second}$, $\Delta\gamma=\gamma^{\rm max}-\gamma^{\rm second}$, as a function of the lattice size $N$ and for some representative values of $\alpha$. For $\alpha <1/2$, the gap increases with system size, with the largest gap for the case for $\alpha = 0$, while for $\alpha >1/2$, the gap closes as the system size increases. The exponent  $\alpha = 1/2$ separates these two regimes. At this value, the gap is independent of $N$. This behavior, depicting a change in the scaling of the imaginary gap as a function of $\alpha$, can  be interpreted as a \textit{``spectral phase transition.''}

Further insight on the short-time behavior for $\alpha<1/2$ can be gained by writing the survival probability as the sum of two contributions:
\begin{equation}
    S(t)\simeq \frac{2}{N}{\rm e}^{-2 (\tau \gamma^{\rm max})t}+2\int_0^\pi\frac{{\rm d}k}{\pi}{\rm e}^{-2 (\tau \gamma_k) t}\,,
    \label{S:t:gap}
\end{equation}
where we have approximated the initial state $|0\rangle\simeq \sum_k |k\rangle/\sqrt{N/2}$ and taken the continuum limit to describe the contribution of all other modes than the mode $k=0$. This separation is justified by the finite gap $\Delta\gamma$ separating the contribution of the first term from the other contributions. Expression \eqref{S:t:gap} shows that, in a finite lattice, there is a finite probability to measure the walker at the target site for very short times $1/\gamma^{\rm max}$. These times can even vanish for $N\to\infty$, predicting events where the detector instantaneously clicks. However, the probability of these events scales with $1/N$ and vanishes in the thermodynamic limit. For large lattices the detection probability at short times then becomes dominated by the second largest eigenvalue $\gamma^{\rm second}$ with the associated timescale $t_{\rm second}\sim 1/\gamma^{\rm second}$ (corresponding to the second largest equipotential line of Fig.\ \ref{fig:phase_Diagram}). For $\alpha\lesssim 1/2$, the rate $\gamma^{\rm second}$ decreases as $\alpha\to 0$ and vanishes in the thermodynamic limit $N\to\infty$. Consequently, the survival probability remains constant or, in other words, the walker does not reach the detector. Equation \eqref{eq:gamma:k:main} shows that this is a consequence of the flat spectrum $\mathcal J_{k>0}$ at $\alpha<1/2$, which in turn is responsible of localization of the walker's wave packet  \cite{Celardo:2016,Santos:2016}. In this regime, therefore, even though the minimal hitting time almost vanishes, the long-range hopping tends to freeze the walker. 

\subsection{Equilibration and relaxation}
\label{sec:regime2}

To discuss the fate of the detection probability of a walker, initially localized at the site $\ket{0}$ and under evolution due to the monitored dynamics, we first consider the case of nearest-neighbor hopping. Recall that the initial state spreads over the periodic lattice in time, and reaches the detector site at short times (set by the Lieb-Robinson bound), following the shortest path connecting the detector site and the initial site. In a finite lattice, either with reflecting boundaries or with periodic boundary conditions, this occurs also at a later time, corresponding to propagation along the other, complementary path. This second timescale is clearly visible in Fig.\ \ref{fig:sprob_slice} corresponding to when the survival probability undergoes a second discontinuous change of the derivative. At this second timescale a process of spreading and interference of the walker's wave packet has set in and continues until equilibration.  The latter would correspond to equal probability of finding the walker at different sites, namely, a uniform distribution over the lattice. Evidently, once equilibrium is attained, all memory of the initial state gets erased.  The occurrence of the second timescale, corresponding to propagation along the complementary path, is evidently a finite-size effect: we have verified that the extent of the steps increases linearly with $N$ for nearest-neighbor hopping. 
 
Using this picture, on a ring lattice geometry the timescale is of the order of the complementary distance $N-D$ between the initial site and the detector, while equilibration will be attained after a time corresponding to several round trips around the lattice, thus over a time that is of the order of multiples of $N$. In Fig.\ \ref{fig:phase_Diagram}, these regimes correspond to different color scales at $\alpha>2$: the transient red-orange region, where the survival probability remains almost constant after the first decay, corresponds to the time elapsed before the walker reaches the detector site along the complementary path. Interference sets in at the second step of the survival probability, which then starts equilibrating towards the asymptotic value. 

If we now turn to long-range hopping, this picture helps us interpret the behavior for $\alpha>1/2$, with some salient differences: For $\alpha\in(1/2,2)$, the timescale of the first detection increases monotonically with $\alpha$, while the size of intermediate region becomes more blurred. At $\alpha\simeq 1/2$, in particular, equilibration sets in at a later time than that for the causal propagation. For $\alpha\in(0,1/2)$ the equilibration timescale scales as $1/\alpha$ and is largest as $\alpha\to 0^+$. In this regime, there is no notion of distance: The walker is either localized or spread across the lattice.  

Fidelity is an important metric that quantifies the attainement of equilibration in a quantum system~\cite{Defenu2021MetastabilitySystems}. It is defined in our setting as
\begin{align}
    f(t) \equiv \|\langle{0}| \psi(t) \rangle \|^2,
    \label{eq:fidelity}
\end{align}
and is thus the norm-squared of the overlap between the initial site $\ket{0}$ and the time-evolved un-normalized state $\ket{\psi(t)}$. For a quantum walk, as is our case, it is the return probability. In the absence of monitoring, one expects that the fidelity will vanish on attaining equilibration, $f(t)\to 0$. 

\begin{figure}[!htpb]
\includegraphics[width=\columnwidth]{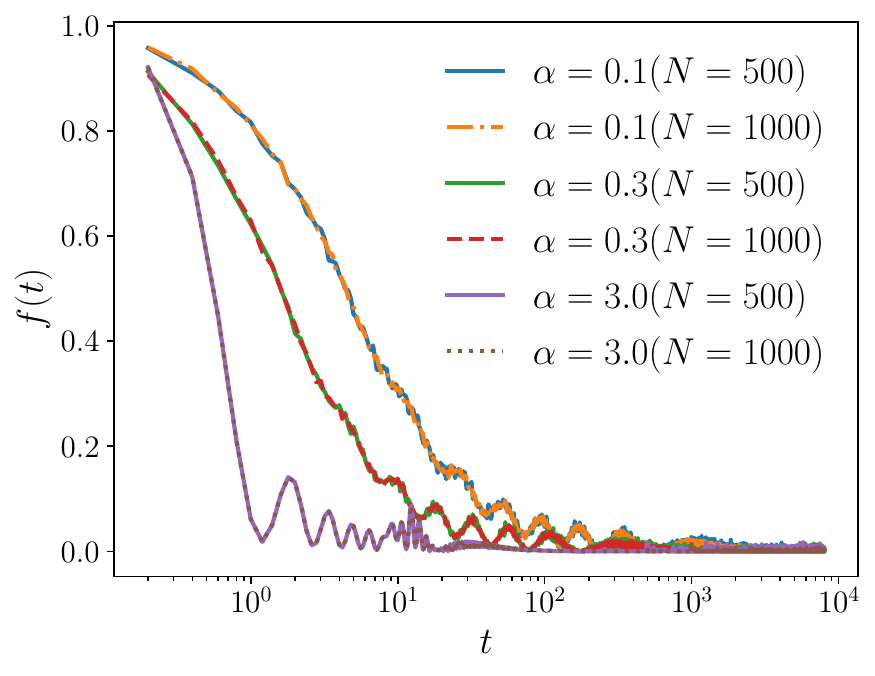}
    \caption{Evolution of the fidelity, Eq.~\eqref{eq:fidelity}, for different values of $\alpha$ and $N$. The parameters are $\tau = 0.2/J,J = 1, D = 10$. The plot does not show the value $f(0)=1$.}
    \label{fig:Equilibration}
\end{figure}

\begin{figure}[!htpb]
\includegraphics[width=\columnwidth]{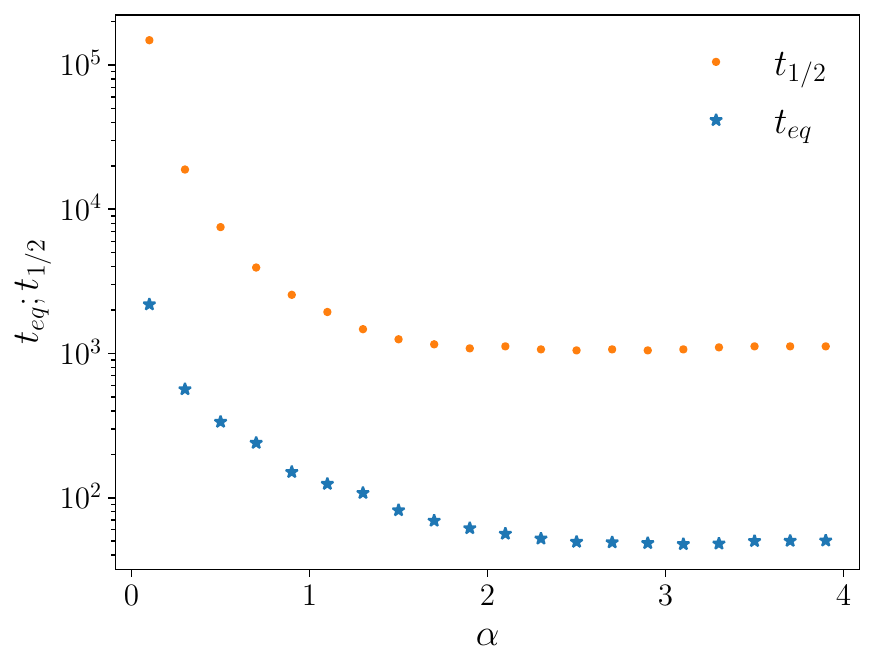}
    \caption{Relaxation and equilibration timescales, respectively $t_{1/2}$ and $t_{\rm eq}$, plotted against $\alpha$. The results are obtained from numerical simulation of the monitored quantum walk, and they correspond to parameter values $N=1000$, $\tau = 0.2/J,J = 1, D = 10$. The equilibration timescale $t_{\rm eq}$ is defined here as the time beyond which $f(t)$ remains at a value smaller than a threshold value set equal to $0.01$. The relaxation timescale is the time at which the survival probability reaches the value $0.75$, thus 50\% between unity and the asymptotic value 1/2.}
    \label{fig:t_eq}
\end{figure}

An approximate expression for the fidelity can be evaluated in the limit of $J\tau\ll 1$, which we have been considering so far. In this limit, we describe the dynamics of the chain using the spectral decomposition of the effective Hamiltonian, and we rewrite the fidelity as interference of waves, which are damped by the rates $\gamma_a$:
\begin{align}
    f(t) = |\sum_a e^{-(i \lambda_a^{(0)} + \tau \gamma_a) t} \, \bra{0}\lambda_a \rangle \langle \bar \lambda_a\ket{0}|^2\,.
    \label{eq:ft-explicit}
\end{align}
This expression predicts oscillations due to frequencies $\lambda_a^{(0)}$, which are damped at rates $\gamma_a\tau$. The dynamics of equilibration results from the interplay between the rates of the incoherent dynamics due to monitoring, and the interference of the free waves propagating across the chain. Figure \ref{fig:Equilibration} displays the evolution of the fidelity, for three representative values of the exponent $\alpha$, where we choose $\alpha=3.0$, representing the case where the dynamics starts to approach the nearest-neighbor limit, and $\alpha<0.5$, where instead localization effects are expected. In all cases, we observe three regimes: an initial decay, oscillations, and an asymptotic regime, where the fidelity tends towards a stationary value close to zero.  With respect to the short-range regime, however, the long-range dynamics slows down the evolution by more than one order of magnitude: even the periodicity of the oscillations is longer at $\alpha=0.3$ and even longer for $\alpha = 0.1$.  

The equilibration time, with which the fidelity decays to zero, is clearly related to the relaxation timescale, which characterizes the decay of the survival probability. Nevertheless, there are salient differences, since the equilibration time results from the interplay of interference and decay. In the regime we consider, where the effect of the detector is a perturbation to the unitary dynamics, we expect that interference is the dominant mechanism leading to equilibration, while relaxation is slower. Figure~{\ref{fig:t_eq}} shows that the relaxation timescale is a multiple of the equilibration one for all values of $\alpha$. Both increase as $\alpha$ decreases, indicating that long-range interactions tend to inhibit the achievement of equilibrium.

\subsection{Universal regime}

\begin{figure}[!htpb]
\includegraphics[width=\columnwidth]{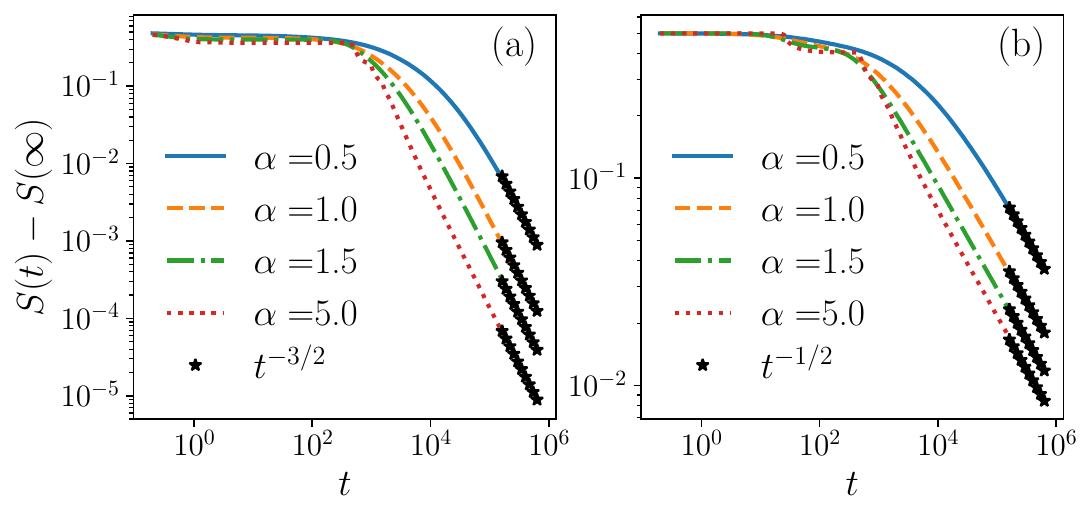}
    \caption{Long-time behavior of the survival probability $S(t)$ for different values of $\alpha$.  Here $N= 1000, \tau = 0.2/J, J=1$, while the distance between the initial location of the walker and the target site is (a) $D = 1$ and (b) $D = 50$. The black star points correspond to a fit with (a) $g(t)\propto t^{-3/2}$ and (b) $g(t)\propto t^{-1/2}$. The time is in units of $1/J$.}
    \label{fig:Universal}
\end{figure}

We now turn to the universal regime. Figure \ref{fig:Universal} shows that the survival probability decays algebraically with time to the asymptotic value with an exponent independent of $\alpha$. We identify two behaviors, which are determined by the distance $D$ between initial site and detector. For distances $D>D^*\gtrsim 1$, the survival probability decays towards the stationary value following the power law $t^{-1/2}$, while for $1\le D<D^*$ the power law is $t^{-3/2}$. This behavior was already reported for nearest-neighbor hopping \cite{Dhar2015DetectionMeasurements,Dhar2015QuantumModel}. Here, we show that it is independent of the hopping range. In fact, as we argue below, it is solely due to the spatial symmetries of the setup, which are independent of the hopping range. Due to the independence of $\alpha$ we denote this regime as universal.

We start by recalling the existence of dark eigenmodes, namely eigenmodes with $\gamma_a=0$. These eigenmodes are $N/2$ and in our model are well approximated by standing waves with a node at the detector site. They are described by standing waves of the form $|k_D\rangle= \sum_j|j\rangle\sin(k(j-D))/\mathcal N$, with $\mathcal N\simeq \sqrt{N/2}$ (see Appendix~\ref{app2}). The standing waves have a node at $D$ . Their wave number takes the value $k=2 \pi n/N$ with $n\in [0,\ldots,N-1]$ and the dark subspace has degeneracy $N/2$. Clearly, the dark eigenmodes are solely due to the geometric properties of the setup. Of these modes, a subset does not contribute to the dynamics: This subset is composed of the modes with wavelength $\lambda=2\pi/k=D/m$ (with $m=1,2,\ldots,D/2$), which also have a node at the initial site. 

We first show how these features determine the asymptotic behavior of the fidelity. In fact, the dark eigenmodes frequencies span the full spectrum $\mathcal J_k$, Eq.\ \eqref{eq:spectrum}, and their interference is responsible for the asymptotic behavior $f(t)\to 0$ for $t\to\infty$ \cite{Defenu2021MetastabilitySystems}. Thus, despite the non-Hermitian dynamics, the dynamics of the fidelity at the asymptotics is determined by the unitary evolution within the dark subspace.

The asymptotic decay of the survival probability can be understood in terms of the eigenmodes that have finite, but small overlap with the target site, which can thus be very long lived. Stretching a definition of quantum optics \cite{Godun:1999}, they are gray eigenmodes. For sufficiently large $t$, one can approximate the expression for the survival probability, Eq.\ \eqref{eq:Sn-2}, as
\begin{eqnarray}
\label{Eq:D}
    S(t)  &\approx& S(\infty)  + \frac{1}{\pi}\int_{-\infty}^{\infty} \mathrm{d}\delta\, e^{- \frac{8 t \tau}{N}  \tilde{J}^2 \delta^2}\sin^2(\delta D) \,,
\end{eqnarray}
where $\tilde{J} $ is the modified prefactor and we have used that only small values of the argument of the exponential contribute at long times. Moreover, we have approximated the decay rates with their functional dependence for the long-wavelength modes ($\delta\ll 1$). In Appendix \ref{app3} we show that the integral leads to the scalings $t^{-1/2}$ and $t^{-3/2}$ for $D>D^*$ and $D=1$, respectively, where the value of $D^*$ depends on the size of the lattice $N$.
The form of integral \eqref{Eq:D} is independent of $\alpha$, as it is dominated by the contribution of the ``gray'' standing waves. The power-law scaling is reminiscent of the Levy-flight statistic reported for the trapping times in gray states of laser cooling \cite{Bardou:1994}.

The return problem, where the initial and detection site coincide ($D=0$), deserves a separate discussion. In this case the behavior cannot be understood in terms of dark eigenmodes; integer exponents have been observed in the long-time tails of survival probability for nearest-neighbor hopping~\cite{Friedman2017QuantumWalk,Friedman2017QuantumProblem, walter2023thermodynamic}. The analysis for long-range hopping will be object of future investigations.

Finally, similar universal long-time behavior is expected to emerge in stroboscopically monitored discrete-time quantum walks (DTQWs) with long-range hopping. In DTQWs~\cite{Shukla:2025}, the additional degree of freedom given by the coin rotation influences the position probability amplitude on lattice sites. Consequently, the position probability distribution is  bipartite in the sense that odd (even) sites are empty at even (odd) detection steps. Because the detection probability is then concentrated on fewer sites, the total (cumulative) detection probability $P_{\mathrm{det}}(n)$ grows faster than in the continuous time quantum walk and also saturates at a larger value, as already observed for nearest-neighbour monitored DTQW~\cite{Shukla:2025,Thiel:2018}. At long times, the dynamics will still be governed by dark (or ``gray”) modes solely determined by the lattice symmetry. Consequently, the universal algebraic decay of the survival probability, and its independence of the hopping exponent $\alpha$, should persist in the DTQW.

\section{Optimal quantum resetting protocol}
\label{Sec:QSR}

Whenever resetting offers an advantage in a classical setting, it has been rigorously demonstrated that among various resetting mechanisms, resetting at a constant pace (sharp-resetting) represents the optimal strategy for minimizing the mean first-passage time~\cite{Chechkin:2018,Pal:2017}.  While alternative protocols (e.g. time-dependent or random reset intervals) can exhibit rich dynamics~\cite{Evans2020StochasticApplications}, their performance is ultimately bounded by that of a properly chosen constant-reset protocol. Subsequent studies in the quantum domain, particularly on monitored quantum walks, have shown that broadening the reset distribution to Poisson or geometric forms not only raises the mean hitting time but also suppresses the quantum-coherent oscillations responsible for ballistic advantage. As a result, sharp-resetting remains the benchmark for optimality even in the quantum domain, although interference effects can create additional, system-specific local minima around the global optimum~\cite{Yin2023RestartTimes}. We now address the open question:  what is  the optimal reset time for our specific quantum system?

In our model, the survival probability analysis singles out a microscopic timescale, set by the hopping exponent~$\alpha$, that helps us design the optimal protocol, thereby accelerating the convergence to the target site. We fix the resetting time to the value $t_R=r\tau$, and discuss the optimal resetting rate $\Gamma_r=1/(r\tau)$ as a function of $\alpha$.

For short-range models we could identify a lowest bound to the hitting time $t_{\rm min}\sim \tau_{\rm LR}$ ($\alpha>2$), such that any resetting faster than $t_{\rm min}$ will impede the walker to reach the target.  The stepwise dynamics of the survival probability, Fig.\ \ref{fig:sprob_slice}, shows that shortly after $t_{\rm min}$ the survival probability reach a transient, constant behavior and suggests therefore that the optimal resetting time shall be $t_R\gtrsim t_{\rm min}$.

This strategy can be extended to the long-range regime by setting the resetting time to the fastest timescale of the dynamics. This implies the choice 
\begin{equation}
\label{eq:r:opt}
      t_R = r_{\rm opt}\tau \sim 1/ \gamma^{\rm max}\,.
\end{equation}
For $\alpha\lesssim 1/2$ the rate $\gamma_{\rm max}$ can become much larger than the inverse of the time interval $\tau$ between two consecutive detections. In this regime, moreover, the probability of supersonic detections scales with $1/N$. This  suggests that for large lattices the resetting rate should scale with the lattice size, $\Gamma_r\sim N$ (thus $r\sim 1/N$), in order to have significant probability of a detection.

\begin{figure}[!htpb]
\includegraphics[width=\columnwidth]{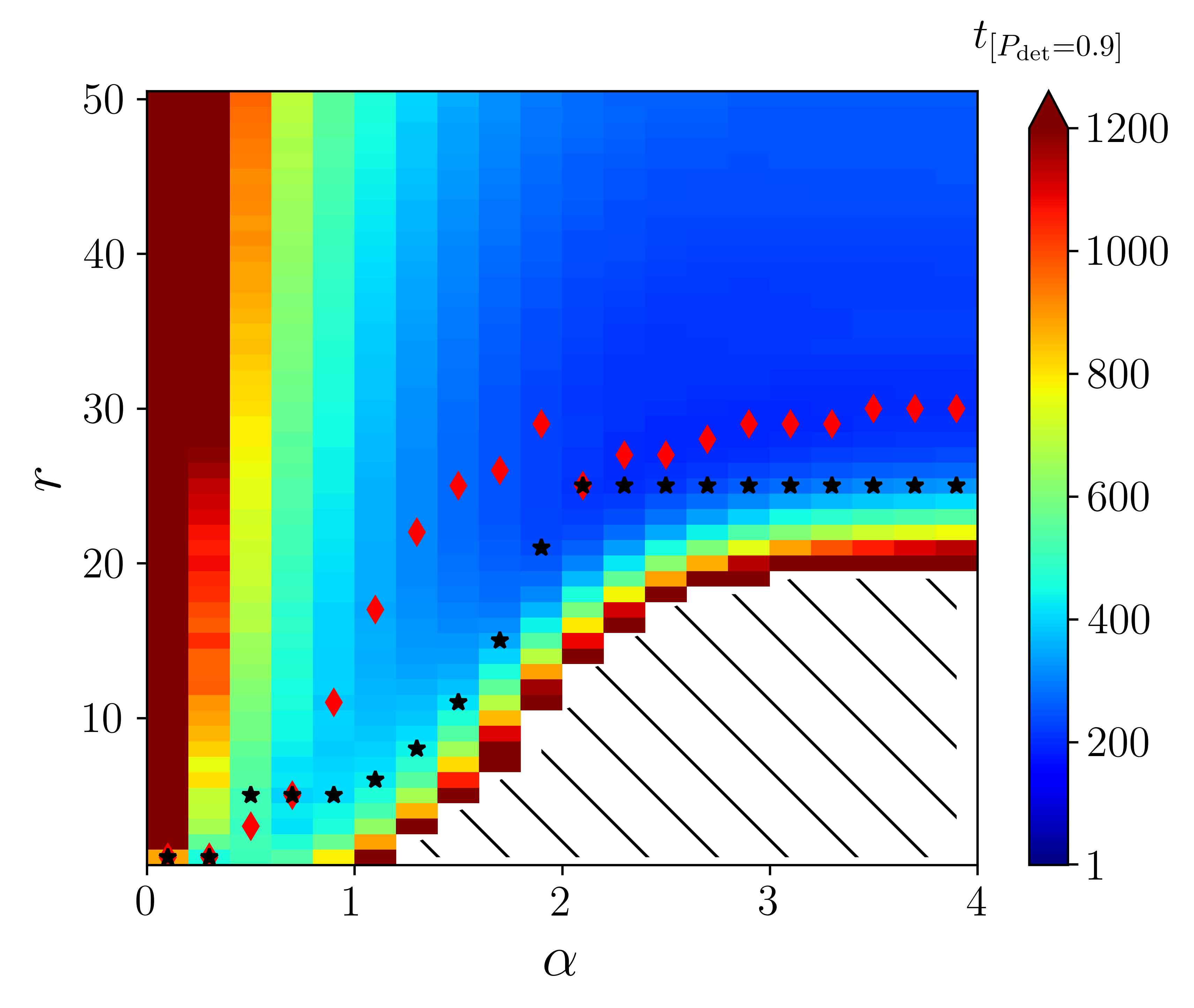}
    \caption{Convergence time of the quantum resetting procotol as a function of $r$ and of the hopping exponent $\alpha$. The convergence time is defined here as the time needed to reach to the detection probability $P_{\rm det}=0.9$ . The black stars correspond to the reset time interval $t_{\rm{R}}=r\tau$ of Eq.\ \eqref{eq:r:opt}, and the red diamonds represent the optimum reset rate found from numerical simulations. In the diagonal hatched region, $P_{\rm det}$ is always zero. The parameters are $N= 1000, \tau = 0.2/J, J=1, D = 10$.}
    \label{fig:reset_opt}
\end{figure}

\begin{figure}[!htpb]
\includegraphics[width=\columnwidth]{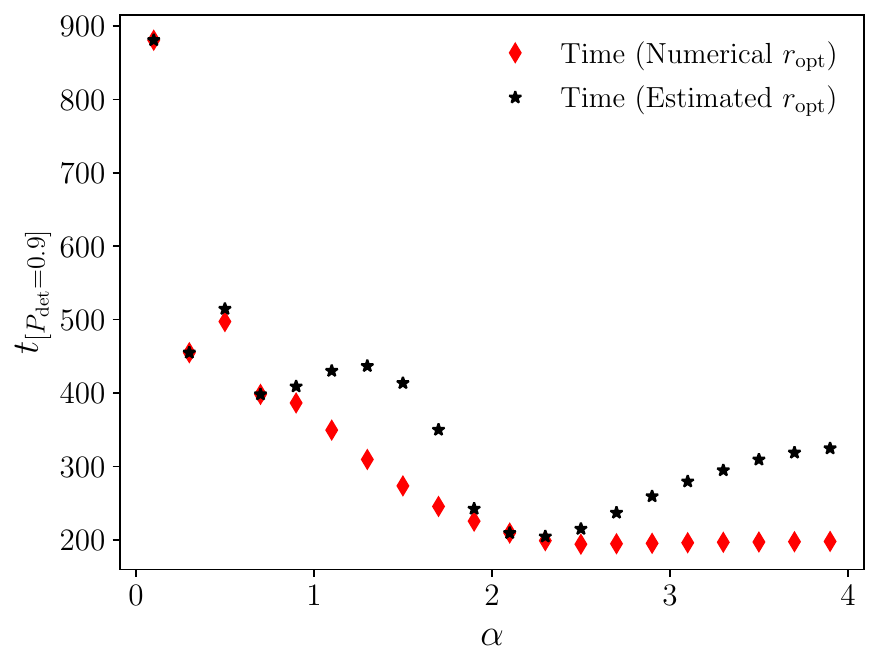}
    \caption{The minimal convergence time $t_{[P_{\rm det} = 0.9]}$ as a function of $\alpha$. The data correspond to the time along the black stared points and the red diamond points in Fig.\ \ref{fig:reset_opt}.}
    \label{fig:time:reset}
\end{figure}

We verify these hypotheses by means of numerical simulations at fixed lattice sizes. However, when $\gamma_{a}^{\rm max}\tau>1$, we set $r=1$ (thus $t_R=\tau$), corresponding to setting an upper bound to the resetting rate. We then determine the timescale of the quantum resetting protocol by means of the time at which the detection probability $P_{\rm det}$ reaches the value of $0.9$. Figure~\ref{fig:reset_opt} shows the characteristic timescale as a function of the parameter $r$ and of the exponent $\alpha$. The color bar represents different values, progressing from blue for smaller $t_{[P_{\rm det} = 0.9]}$ to red for larger $t_{[P_{\rm det} = 0.9]}$. The black stars show the predictions of the protocol using the reset $r$ of Eq.\ \eqref{eq:r:opt}, while the red diamonds correspond to the optimal reset $r$ determined using numerical optimization, minimizing $t_{[P_{\rm det} = 0.9]}$ at fixed $\alpha$. 

Figure \ref{fig:time:reset} displays the minimal detection time as a function of the hopping exponent. For our implementation, where the minimal resetting time cannot be smaller than $\tau$ and hence  $r\ge 1$, long-range hopping events do not speed up the quantum resetting protocol: the fastest convergence is achieved for hopping exponents $\alpha>1$. This perhaps counterintuitive conclusion is a manifestation of the long-range induced localization that we have analyzed in the survival probability dynamics.


\section{Conclusions}
\label{Sec:Conclusions}

In this work, we analyzed the hitting time of a quantum walk in a one-dimensional lattice. The dynamics alternates coherent evolutions with detection events, modeled by a projection onto the target site. We determined the dynamics of the detection probability for diffusion processes, where the probability amplitude of tunneling decays as a power law $\sim d^{-\alpha}$ with the Euclidean distance $d$ between any two sites of the lattice. Although resembles Lévy flights, this setup does not have a one-to-one classical counterpart, and thus our results advance first-passage studies in quantum walk based search protocols beyond classical analogs. We established an explicit connection between the detection probability and the Lieb-Robinson time at which correlations between distant sites are established by the coherent dynamics, and we showed that the Lieb-Robinson time determines the short-time dynamics of the detection probability, while the dynamics at later times results from the interplay between the coherent propagation and the incoherent dynamics induced by monitoring. This permits us to determine the optimal resetting rate by setting the resetting time to the fastest timescale of the dynamics. Interestingly, our results link the local measurement outcome to the global evolution of the system. 

The dynamics considered here can provide important information, as well as guidance for studying many-body dynamics in the presence of dissipation, such as in Refs.\ \cite{Kessler:2012,Cai:2013,Poletti:2013,Sciolla:2015,walter2023thermodynamic}. In first place, this is possible by connecting the insights on the short-time dynamics to the speed limits of many-body systems \cite{Chen:2023}. At longer times, the evolution of the detection probability in the absence of resetting sheds light on the equilibration and relaxation mechanisms. In the case of a single excitation, the two can be distinguished. Relaxation mechanisms, in particular, set in on timescales determined by the gap between the imaginary eigenvalues. Moreover, due to the presence of eigenmodes that are perfectly decoupled from the detector, the asymptotic dynamics is reduced to a coherent evolution in the subspace spanned by the dark eigenmodes and it is independent of the interaction range. Similar features can be identified in many-body systems on lattices with variable interaction strength~\cite{Sciolla:2015,Roy:2023,Halati:2025}. 

Our study sets the basis for determining the scaling of the hitting time for arbitrary distances in the lattice of size $N$ and as a function of $\alpha$. Our analysis assumes parameters for which the monitoring events occur at stroboscopic timescales that can be well approximated by a continuum. Future work shall identify the dynamics of long-range hopping when the stroboscopic nature of the monitoring events becomes relevant. In this regime, topological features have been predicted for nearest-neighbor hopping \cite{Liu:2023} which are expected to be modified or even suppressed by long-range diffusion. 

Quantum stochastic resetting is a realization of a quantum randomized algorithm. This can be considered the randomized counterpart of a quantum walk, performing a unitary spatial search on a graph \cite{Childs:2004}. Preliminary studies indicate a definite advantage of searches based on monitored quantum walks with respect to coherent dynamics for certain graph structures~\cite{Nzongani:2025}. Understanding whether quantum stochastic resetting might provide an advantage requires one to determine the time complexity of the quantum stochastic resetting algorithm, such as the scaling of the convergence time as a function of the lattice size, that shall be performed after identifying the optimal conditions for a given geometry and diffusion matrix. However, while optimization in the classical counterpart has well-established paradigms in computer science \cite{Luby:1993}, in the quantum regime there is no established framework. Our work is a first step in this direction thanks to the systematic characterization of the timescales as a function of the physical parameters. The development of an optimization framework for the incoherent dynamics shall extend the concepts of Refs.~\cite{King:2024,King:2025} to these settings, and it will be the object of future work.

Finally, our predictions could be verified by existing experimental setups, for instance setups with Rydberg,  dipolar, ion arrays, or more general to settings where the size of the lattice can be varied and the hopping exponent can be tuned from global to nearest-neighbor hopping~\cite{Periwal:2021,Kotibhaskar:2024}. The monitoring procedure can be implemented by electron shelving at the target site, while the restarting procedure requires reinitialization of the quantum system and is a classical feedback mechanism~\cite{Wiseman:1994,Candeloro:2022}.

\acknowledgements The authors acknowledge discussions and helpful comments from Francesco Mattioti, Emma King, Gabriele Perfetto, and Markus Bl\"aser. This work was funded by the Deutsche Forschungsgemeinschaft (DFG, German Research Foundation) – Project-ID 429529648 – TRR 306 QuCoLiMa (``Quantum Cooperativity of Light and Matter''), and the QuantERA project ``QNet: Quantum transport, metastability, and neuromorphic applications in Quantum Networks'' - Project ID 532771420. S.G. acknowledges Alexander von Humboldt Foundation for support under its Humboldt Research Fellowship for Experienced Researchers. He also thanks ICTP–Abdus Salam International Centre for Theoretical Physics, Trieste, Italy, for support under its Regular Associateship scheme. Part of this research was performed at the Kavli Institute for Theoretical Physics (KITP) in Santa Barbara and was supported by Grant No. NSF PHY-2309135 to the Kavli Institute for Theoretical Physics (KITP).

\section*{Data Availability}

The data that support the findings of this article are openly available~\cite{roy_2025_data}.

\appendix

\section{Derivation of the non-Hermitian Hamiltonian}
\label{app1}
Here, we derive Eq.~\eqref{eq:Sn-2} of the main text. To this end, we introduce the projector $\hat P_D=|D\rangle\langle D|$ and split the modified time-evolution operator $\hat{\widetilde{U}}$ in Eq.~\eqref{eq:S_n} in terms of its contribution from the detector subspace and from its complementary subspace by using the fact that the corresponding projection operators satisfy $\hat{P}_D+\hat{P}_D^c=\hat{I}_N$. We have
\begin{align}
    \hat{\widetilde{U}} &= \hat{P}_D^c \hat{U} = \hat{P}_D^c \left[ (\hat{P}_D^c+\hat{P}_D) \hat{U} (\hat{P}_D^c +\hat{P}_D)\right] \nonumber \\
    &= \hat{P}_D^c \hat{U} \hat{P}_D^c + \hat{P}_D^c \hat{U} \hat{P}_D\,,
\end{align}
where we have used $(\hat{P}^c_D)^2=\hat{P}^c_D$ and $\hat{P}^c_D \hat{P}_D=0$. In the same manner, the initial state may be rewritten as 
\begin{align}
    \ket{\psi_0} =  \hat{P}_D^c  \ket{\psi_0} + \hat{P}_D \ket{\psi_0}\,.
\end{align}
As mentioned in the main text, we always choose the initial state to be not in the detector space (i.e., initially, the particle could be on any site other than the target site $D$), which implies that the second term on the right hand side vanishes, and hence the initial state is simply
\begin{align}
    \ket{\psi_0} =  \hat{P}_D^c  \ket{\psi_0}.
\end{align}
The un-normalized state $\ket{\psi_n^+}$ at the end of time $n\tau$ and
just after the $n$-th measurement then has the form
\begin{align}
   (\hat{\widetilde{U}})^n   \ket{\psi_0} &=   [\hat{P}_D^c \hat{U} \hat{P}_D^c + \hat{P}_D^c \hat{U} \hat{P}_D]^n \hat{P}_D^c  \ket{\psi_0} \nonumber \\
   &= [\hat{P}_D^c \hat{U} \hat{P}_D^c]^n  \ket{\psi_0}\,,
   \label{eq:U-3}
\end{align}
which leads to the result
\begin{align}
\hat{\widetilde{U}}=\hat{P}_D^c \hat{U} \hat{P}_D^c\,.
\label{eq:U-tilde-1}
\end{align}
Due to the fact that $\hat{P}_D^c|D\rangle=0$, it follows from the above equation that the effective evolution operator $\hat{\widetilde{U}}$ has one trivial eigenvalue equal to zero, with the corresponding eigenvector given by $|D\rangle$. The remaining eigenvectors, which evidently lie in $\mathcal{H}_D^c$, obviously have zero components in the detector Hilbert space $\mathcal{H}_D$. 

To proceed, we invoke a perturbative-theory treatment justified on the grounds that $\tau$ is small,  i.e., $\tau \ll 1/J$~\cite{Dhar2015QuantumModel,Dhar2015DetectionMeasurements}. One finds straightforwardly on retaining terms up to $O(\tau^2)$ that $\hat{\widetilde{U}}$ in Eq.~\eqref{eq:U-tilde-1} reads 
\begin{align}
    \widetilde{\hat{U}}
    &=e^{-i\hat{H}_\mathrm{eff}\tau};~\hat{H}_\mathrm{eff}\equiv \hat{P}_D^c \hat{H} \hat{P}_D^c -\tau \left( \frac{i}{2} \hat{\Gamma} \right)\,,
    \label{eq:U-2}
\end{align}
with the operator $\hat{\Gamma}$  defined as
\begin{align}
    \hat{\Gamma}\equiv (\hat{P}_D^{c})^2\hat{H}^2\hat{P}_D^{c}-(\hat{P}_D^c\hat{H})^2\hat{P}_D^c =\hat{P}_D^{c} \hat{H}|D\rangle\langle D| \hat{H}\hat{P}_D^c\,.
\end{align}
Note that $\hat{P}_D^c \hat{H} \hat{P}_D^c$ is symmetric and Hermitian, and is obtained from $\hat{H}$ by replacing in the corresponding matrix all entries in the row and the column that refer to the detector site with zero. Moreover, $\hat{\Gamma}$ is also symmetric and Hermitian. Therefore, 
the non-Hermitian Hamiltonian $\hat{H}_\mathrm{eff}$, Eq.~\eqref{eq:U-2}
is symmetric.

\section{Spectrum of the non-Hermitian Hamiltonian}
\label{app1:A}
Let us now obtain the eigenvalues and eigenvectors of $\hat{H}_\mathrm{eff}$ in Eq.~\eqref{eq:U-2}. Considering that the small parameter $J\tau\ll 1$ scales operator $\hat\Gamma$, one may invoke time-independent perturbation theory~\cite{sakurai} and treat the operator $\hat{H}_0\equiv\hat{P}_D^c \hat{H}\hat{P}_D^c $ as the bare Hamiltonian and the operator $\hat V=-{\rm i}\tau\hat{\Gamma}/2$ as a perturbation. Both operators are defined in the Hilbert space $\mathcal H_c$ consisting of all the lattice sites excluding the detector site, such that dim$\mathcal H_c=N-1$. Let $|\lambda_a^{(0)}\rangle$ be the eigenvectors of $\hat H_0$ satisfying the eigenvalue equation
\begin{align}
 \hat{H}_0 |\lambda_a^{(0)}\rangle = \lambda_a^{(0)} |\lambda_a^{(0)}\rangle\,,
 \label{eq:eigenvectors}
\end{align}
with $\langle \lambda_a^{(0)}|\lambda_{a'}^{(0)}\rangle=\delta_{a,{a'}}$. The eigenvalues $\lambda_a^{(0)}$ form the spectrum of a Hermitian operator and are therefore real. The corrections $\lambda_a^{(1)}$ are calculated in first order perturbation theory. In first order, the corrections, $\lambda_a^{(1)}=\langle\lambda_a^{(0)}|\hat V|\lambda_a^{(0)}\rangle$, are imaginary. For convenience, we write $\lambda_a^{(1)}=-{\rm i}\gamma_a\tau$, where $\gamma_a$ is real and positive:
\begin{align}
    \gamma_a = \frac{1}{2}  |\langle\lambda_a^{(0)}|\hat H |D\rangle|^2\,, 
    \label{eq:gamma_a}
\end{align}
where we used the explicit form of $\hat V$ and recall that $\hat H$ is Hamiltonian \eqref{eq:Hamiltonian}. Replacing the explicit form of $\hat H$, the above equation can be recast into the form 
\begin{align}
    \gamma_a = \frac{J^2}{2}  \left|\sum_{i\neq D}\frac{\langle\lambda_a^{(0)}|i\rangle}{d_{iD}^\alpha}\right|^2\,, 
    \label{eq:gamma_a2}
\end{align}
with $d_{i,D}$ the minimal distance between site $i$ and site $D$ along the ring.
Accordingly, the first-order correction to the eigenvectors is given by $\tau \ket{\lambda_a^{(1)}}$, with
\begin{align}
   &\ket{\lambda_a^{(1)}} \equiv - \frac{\rm i}{2} \sum_{b \neq a} |\lambda_b^{(0)}\rangle\frac{ \langle\lambda_b^{(0)}|\hat\Gamma| \lambda_a^{(0)}\rangle}{\lambda_a^{(0)} - \lambda_b^{(0)}}  \,.
\end{align}
Assuming that the walker is initially localized on the site $|0\rangle$ of the lattice, in the first-order perturbation theory, the survival probability takes the form 
\begin{align}
S(t)= \sum_{a}^{} e^{-2 (\tau \gamma_a) t}|\langle\lambda_a^{(0)}|0\rangle|^2 \,.
\label{eq:Sn-app2}
\end{align}

\subsection{Spectral properties in the long-wavelength limit}

\begin{figure}[!htpb]
\includegraphics[width=\columnwidth]{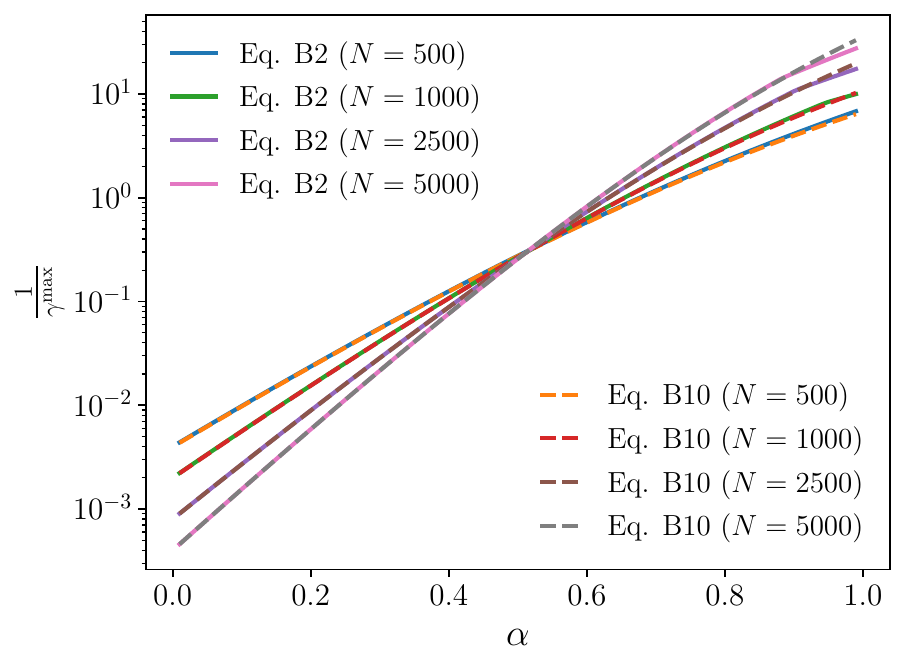}
    \caption{$\gamma^{\rm max}$ is numerically evaluated using Eqs.~\ref{eq:gamma:max} and \ref{eq:gamma_a}, and they are in perfect agreement. Other parameter values are  $\tau = 0.2/J, J=1$, $D =10$.}
    \label{fig:gamma_max_approx}
\end{figure}

\begin{figure}[!htpb]
\includegraphics[width=\columnwidth]{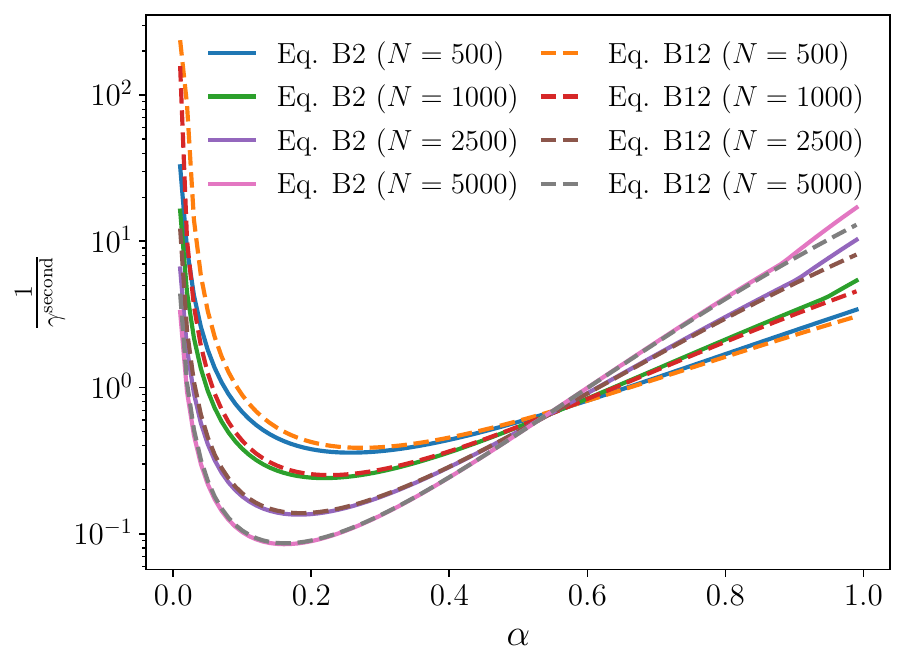}
    \caption{$\gamma^{\rm second}$ is numerically evaluated using Eqs.~\ref{eq:gamma:second} and \ref{eq:gamma_a}, and they are in good agreement. Other parameter values are  $\tau = 0.2/J, J=1$, $D=10$.}
    \label{fig:gamma_second_approx}
\end{figure}

We now derive the explicit form of $\gamma_a$ in some approximate limit. Consider the Hamiltonian $\hat H_0$, corresponding to the original Hamiltonian except for the site $|D\rangle$. In the subspace $\mathcal H_c$,  it is then the Hamiltonian of a lattice where all sites are uniformly spaced, e.g., at distance $d$, except for two neighboring sites that are at distance $2d$, and which correspond to the sites
$|D-1\rangle$ and $|D+1\rangle$ of the original lattice, which are at distance $2d$. For a sufficiently large lattice, $N\gg 1$, the long-wavelength eigenmodes of $\hat{H}$ are well approximated by sinusoidals with wave number $k= 2\pi n/Nd$. This approximation is good for $kD\ll 1$, while the presence of the defect significantly modifies the spectrum at $kD\lesssim 1$. For the moment, however, we treat the effect of the defect as a small correction and assume $\hat H_0$ to be invariant under discrete translation. We then get $|\lambda_a^{(0)}\rangle=|k\rangle$ with 
\begin{equation}
    \langle i|\lambda_a^{(0)} \rangle\simeq \langle i|k \rangle =  \sqrt{\frac{1}{N-1}}\cos(ki)\,,
\end{equation} 
(we do not discuss the $\sin$ waves, since they have no overlap with the initial site $|0\rangle$) and the eigenvalues $\mathcal J_k$, such that $\hat H_0|k\rangle=\mathcal J_k|k\rangle$, with
\begin{equation}
\mathcal J_k=-2J\sum_{r=1}^{N/2 - 1}\frac{\cos(kr)}{r^\alpha}\,,
\label{eq:J_k}
\end{equation}
for even $N$. Using these expressions in Eq.\ \eqref{eq:gamma_a}, we find
\begin{equation}
   \gamma_a \simeq\gamma_k = \frac{1}{N}\mathcal J_k^2 \cos^2(kD)\,,
    \label{eq:gamma_k}
\end{equation}
where we have approximated $N-1\simeq N$. Numerically, it is shown in Figs.~\ref{fig:gamma_max_approx} and~\ref{fig:gamma_second_approx}, respectively, that Eqs.~\ref{eq:gamma:max} and~\ref{eq:gamma:second} are very good approximation of Eq.~\ref{eq:gamma_a}. Correspondingly, using that $|0\rangle\simeq \sum_{k=0}^{N/2} |k\rangle/\sqrt{N/2}$, the survival probability takes the form
\begin{equation}
    S(t)\simeq \frac{1}{N}\sum_{k=0}^{N/2}\exp(-2\tau\gamma_k t)\,.
\end{equation}

\subsubsection{Exponents $0<\alpha<1$}

Within this treatment, for $\alpha<1$ the largest decay rate is at $k=0$, namely $\gamma^{\rm max}=\gamma_{k=0}$, and specifically,
\begin{equation}
    \gamma^{\rm max}\simeq 2J^2\frac{1}{N}\zeta_N(\alpha)^2\,,
    \label{eq:gamma:max}
\end{equation}
with $\zeta_N(\alpha)=\sum_{r=1}^{N/2}1/r^\alpha$, such that in the thermodynamic limit it tends to Riemann's zeta function. This shows that the approximation at the basis of this calculation is reliable for the short-time dynamics, which is dominated by the long-wavelength eigenmodes. 

The function $\zeta_N(\alpha)^2$ is monotonically decreasing with $\alpha$, thus $\gamma^{\rm max}$ decreases as $\alpha$ increases, as we also verify numerically in Fig.\ \ref{fig:gamma_max}. The scaling with $N$ is determined by the ratio $\zeta_N(\alpha)^2/N$. For $\alpha<1$ the function scales as $\zeta_N(\alpha)\sim N^{1-\alpha}$, and $\gamma_{\rm max}\sim N^{1-2\alpha}$. This approximate treatment thus reproduces the behavior observed for $\alpha<1$ in Fig.\ \ref{fig:gamma_max}.

For $0<\alpha<1/2$ the rate $\gamma^{\rm max}$ is separated by a finite gap from all other rates 
$\gamma_k$ with $k>0$, which in turn form a continuum. Therefore
\begin{equation}
    S(t)\simeq\frac{{\rm e}^{-2 (\tau \gamma_{\rm max})t}}{N}+\int_0^\pi\frac{{\rm d}k}{\pi}{\rm e}^{-2 (\tau\gamma_k)t}\,.
\end{equation}
We estimate the scaling of the gap by determining the scaling of the next largest imaginary eigenvalue $\gamma_a^{\rm second}$. This eigenvalue is at $k= 2\pi/N$
\begin{equation}
    \gamma_a^{\rm second} \sim 2J^2 \frac{1}{N} \left(\sum_{r=1}^{N/2 - 1}\frac{\cos(\frac{2\pi r}{N})}{r^\alpha}\right)^2 \,.
    \label{eq:gamma:second}
\end{equation}

\subsubsection{Exponent $\alpha>1$}

For $\alpha>1$ the description in terms of sinusoidal starts to fail: the maximum of $\gamma^{\rm max}$ is for wave length of the order of $D$, where the influence of the defect on the dynamics of the unperturbed states become important. The defect in Hamiltonian $\hat H_0$ couples several sites and leads to a mixing of wave numbers $k$ over a range that decreases with $\alpha$.




\section{Dark states and behavior of the asymptotic survival probability $S(\infty)$}
\label{app2}

The dark states $\ket{\delta}$ are defined as states that are unaffected by our dynamical scheme of unitary evolution interspersed with measurements. Two types of dark states are known to exist: the first type includes so-called Floquet dark states that arise on performing measurements periodically at regular time intervals of duration $\tau$, while the second type involves generic dark states that are a result of non-periodic measurements~\cite{Thiel2020DarkDetection}. One can systematically get rid of the Floquet dark states by appropriately choosing the measurement interval $\tau$. We discuss here the generic dark states that arise due to degeneracy in the eigenspectrum of the Hamiltonian~\eqref{eq:Hamiltonian}. To this end, let us first obtain the eigenspectrum. The Hamiltonian in matrix form reads 

 \begin{align}
     \hat{H} =  \begin{bmatrix}
         v_0 & v_{N-1} &\ldots& v_2 & v_1 \\
         v_1 & v_0 & v_{N-1} & & v_2 \\
         \vdots& v_1 & v_0 &\ddots & \vdots \\
         v_{N-2} & &\ddots &\ddots & v_{N-1} \\
         v_{N-1} & v_{N -2} & \ldots & v_1 & v_0
     \end{bmatrix},
 \end{align}
with $v_0 = 0$ and 
\begin{align}
    v_d \equiv \begin{cases}
        1/d^\alpha \hspace{0.8 cm}\text{if}\,\,\, 1 \leq d \leq N/2 - 1, \\
        1/ (N - d)^\alpha \hspace{0.5 cm} \text{otherwise.}
    \end{cases}
\end{align}
Clearly, $\hat{H}$ is a circulant matrix, which may be expressed as
\begin{align}
    \hat{H} = v_0 \hat{I}_N + v_1 \hat{P} +v_2 \hat{P}^2 + \ldots +v_{N-1} \hat{P}^{N -1},
\end{align}
where $\hat{P}$ is an $N \times N$ cyclic permutation matrix, given by
 \begin{align}
    \hat{P} =  \begin{bmatrix}
        0 & 0 &\ldots& 0 & 1 \\
         1 & 0 & \ldots & 0 & 0 \\
         \vdots& \ddots  & \ddots  &\vdots & \vdots \\
          0 &\ddots  &\ddots &0 & 0 \\
         0 & 0 & \ldots & 1 & 0
     \end{bmatrix}.
 \end{align}

Now, since $\hat{P}^N = \hat{I}_N$, the eigenvalues of $\hat{P}$ are given by $\omega_a = e^{i 2\pi a/N};~a= 0,1,\ldots N-1$, where $\omega_a$ is the $N$th root of unity. This implies that the eigenvalues of $\hat{H}$ are given by \cite{Barre2002MicrocanonicalInteractions, Defenu2023Long-rangeSystems}
\begin{align}
    \lambda_a = \sum_{d = 0}^{N-1} v_d \omega_a^{d} = 2 \sum_{d = 1}^{N/2 -1} \frac{\cos(2\pi a d/N)}{d^\alpha} \,\, ;\,\,\,\, a =0,1,\ldots, N-1. 
\end{align}

We will discuss below the nature of the spectrum for different values of $\alpha$.

\begin{itemize}
    \item \textit{Case 1:} $\alpha \neq 0$. 
Except for the ground state $(a =0)$ which corresponds to the eigenvector $1/\sqrt{N} [1,1,\ldots,1]^T$, each of the remaining $N-1$ eigenvalues is twofold-degenerate with the corresponding two eigenvectors given by $\ket{e^a}, \ket{f^a}$; the components of the respective eigenvectors are given by 
\begin{align}
    e_j^a  = \sqrt{\frac{2}{N}} \cos \left( \frac{2 \pi a j}{N} \right);~ f_j^a  =  \sqrt{\frac{2}{N}}  \sin \left( \frac{2 \pi a j}{N} \right), 
\end{align}
with $j = 0,1,\ldots, N-1$. Because of the above-mentioned degeneracy in the spectrum, one can construct dark states through a superposition of the degenerate eigenvectors with proper weights for every $a=1,2,\ldots,N-1$, as follows:
\begin{align}
    \ket{\delta^a} &=    \left(\bra{D}e^a \rangle \ket{f^a} - \bra{D}f^a \rangle \ket{e^a} \right) \nonumber\\
    &= \sqrt{\frac{2}{N}}   \left(  \cos \left( \frac{2 \pi a D}{N} \right) \ket{f^a} -\sin \left( \frac{2 \pi a D}{N} \right) \ket{e^a}\right),
    \label{eq:dark}
\end{align}
such that $\hat{\widetilde{U}}\ket{\delta^a} = \ket{\delta^a}$. The asymptotic survival probability $S(\infty)$ will be due to  states that are not affected by measurements, and it is  thus given by the overlap between the initial state and all the dark states, as
\begin{align}
    S(\infty) &= \frac{2}{N}  \sum_{a \neq 0} \,\, |\bra{l} \delta^a \rangle|^2 \nonumber \\ &= \frac{2}{N}\sum_{a \neq 0}| \cos \frac{2 \pi a D}{N} \sin  \frac{2 \pi a l}{N} -  \sin \frac{2 \pi a D}{N} \cos  \frac{2 \pi a l}{N}|^2 \nonumber \\ &= \frac{2}{N} \sum_{a \neq 0} \sin^2 \left(\frac{2 \pi a (D - l)}{N}\right). 
    \label{eq:S_infty}
\end{align}
In the thermodynamic limit (i.e., $N \to \infty$), one gets
\begin{align}
    S(\infty) = \begin{cases}
        0 \hspace{0.8 cm}\text{if}\,\,\,  D - l = N/2,\\
        1/2 \hspace{0.5 cm} \text{otherwise.}
    \end{cases}
\end{align}

From Eq.~\eqref{eq:dark}, it follows that for each pair of degenerate eigenvectors, one can thus identify a corresponding dark state, leading to a ratio of the number of dark states to the total number of states of $0.5$. In the perturbation-theory approach discussed in Appendix~\ref{app1}, dark states correspond to those modes that do not decay ($\gamma_a = 0$) and the density $n_a/N$ of these modes equals $0.5$.


\item \textit{Case 2:} $\alpha = 0$. This is a special case; here, the spectrum consists of the ground state and an excited state that is $(N-1)$-fold degenerate. In this case, it is easier to write the bright states~\cite{Thiel2020DarkDetection}, i.e., the states that are complementary to the dark states and therefore will always get detected with probability unity. The bright states are obtained as
\begin{align}
    \ket{\beta} =  \sum_{a\neq 0} \ket{e^a} \sin \frac{2 \pi a D}{N} + \ket{f^a} \cos \frac{2 \pi a D}{N}.
\end{align}
The asymptotic total detection probability $P_{\rm det}(\infty)$, which is the complement of the survival probability, can then be described by the overlap of the initial state with the bright states, as
\begin{align}
P_{\rm det}(\infty) &= |\langle l | \beta \rangle|^2 \nonumber \\ &= \frac{2}{N} |\sum_a \langle l | e^a \rangle \sin \frac{2 \pi a D}{N} + \langle l | f^a \rangle  \cos \frac{2 \pi a D}{N}|^2 \nonumber \\  &= \frac{2}{N}  |\sum_a  \sin \frac{2 \pi a l}{N} \sin \frac{2 \pi a D}{N} + \cos \frac{2 \pi a l}{N} \cos \frac{2 \pi a D}{N}|^2 \nonumber \\  &= \frac{2}{N}  |\sum_a \cos \frac{2 \pi a (D-l)}{N}|^2 \to_{N \to \infty} 0.
\end{align}
\end{itemize}
Thus, in the thermodynamic limit, one gets $S(\infty) = 1 - P_{\rm det}(\infty)  \to 1$.

\section{Universal Regime of Survival Probability}
\label{app3}

Using Eqs.~\eqref{eq:Sn-app2} and \eqref{eq:gamma_k}, one can write the survival probability as
\begin{equation}
  S(t) \simeq \sum_{k} e^{- \frac{2 t \tau}{N}\mathcal{J}_k^2\cos^2(kD) }|\langle k |\psi(0) \rangle|^2 \,,
\end{equation}
where $\mathcal J_k$ is given by Eq.~\eqref{eq:J_k}. 

We first review the nearest-neighbor case ($\alpha \to \infty$). In Ref. \cite{Dhar2015DetectionMeasurements} the authors set the detection site at $N$ and the initial site at $l$. This setup is equivalent to our model for $\alpha \to \infty$ and with $D = \min \{ l, N - l\}$. In this setup, the survival probability takes the form
\begin{eqnarray}
      S(t) &\simeq& \frac{2}{N}\sum_{k} e^{- \frac{8 t \tau}{N} J^2 \cos^2(k)}\cos^2(k l)+ S(\infty)\,, \nonumber \\
\end{eqnarray}
 where $S(\infty ) = \frac{2}{N}\sum_{k} \sin^2(kl)$. In the thermodynamic limit ($N \to  \infty$)$, S(\infty)$ matches Eq.~\eqref{eq:S_infty}, yielding either $1/2$ or $0$ depending on the initial site position. The first term of the above equation can be converted in the integral form  as
\begin{equation}
      S(t)- S(\infty) \approx \frac{2}{\pi}\int_0^\pi \mathrm{d}k e^{- \frac{8 t \tau}{N} J^2 \cos^2(k)}\cos^2(kl)\,.
\end{equation}

For sufficiently large $t$ ($t\gg N\tau$), the function $e^{-g \cos^2(k)}$ is sharply peaked at $k = \pi/2$. These are the points where $\cos(k) = 0$, meaning away from these points, the function quickly decays. To approximate the integral, we first expand $\cos^2(k)$ around $\pi/2$. Let us consider $k = \pi/2 + \delta$, where $\delta$ is small. Then $\cos^2{k} \approx \delta^2$, and $\cos^2(kl) = \cos^2((\pi/2 + \delta)l)$. The integral will now be approximated as

 \begin{eqnarray}
       S(t)- S(\infty) &\approx& \frac{1}{\pi}\int_{-\infty}^{\infty} \mathrm{d}\delta \,\, e^{- \frac{8 t \tau}{N} J^2 \delta^2}\cos^2((\frac{\pi}{2} + \delta)l) \nonumber \\ 
       &\approx&\frac{1}{\pi} \frac{\sqrt{\pi}}{2J \sqrt{8 t \tau /N}} \left(1  + (-1)^l e^{-\frac{l^2}{(8 J^2 t \tau/N)}} \right). \nonumber \\
       \label{eq:universal}
 \end{eqnarray}

This expression exhibits two distinct power-law decay behaviors, depending on the distance between the initial and detection sites:
\begin{enumerate}[(i)]
    \item For distance less than $D^*$: In this regime, one may Taylor expand the second term of Eq.~\eqref{eq:universal}, and the survival probability decays as $t^{-3/2}$, and then for sufficiently large times, it decays exponentially. $D*$ is determined by the validation of the Taylor expansion such that for distances $D < D^*$, it satisfies $\frac{N D^2}{(8 J^2 t \tau )} \ll 1$.
    \item For distance greater than $D^*$: One can now neglect the second term of Eq.~\eqref{eq:universal} and the decay follows $t^{-1/2}$ decay and then for large enough times decays exponentially.
 \end{enumerate}
These scaling behaviors are consistent with the results in \cite{Dhar2015DetectionMeasurements, Dhar2015QuantumModel}. For the long-range case, only the eigenvalues are modified, but the structure of the integral remains unchanged, as shown in Eq.~\eqref{Eq:D}. Consequently, the same two scaling regimes arise as in the nearest neighbor case (see Fig.~\ref{fig:Universal}).


\bibliography{references}

\end{document}